\documentclass{article}
\usepackage{amsmath,amsfonts,amssymb}
\usepackage{graphicx}
\usepackage{setspace}
\usepackage{tocloft}
\usepackage{cite}
\usepackage{authblk}
\usepackage[utf8x]{inputenc}
\usepackage{lmodern,textcomp}
\usepackage{algorithmic}
\usepackage{textcomp}
\usepackage{colortbl}
\usepackage{xcolor}

\usepackage{PRIMEarxiv}

\usepackage[T1]{fontenc}    
\usepackage{hyperref}       
\usepackage{url}            
\usepackage{booktabs}       
\usepackage{amsfonts}       
\usepackage{nicefrac}       
\usepackage{microtype}      
\usepackage{lipsum}
\usepackage{fancyhdr}       
\graphicspath{{media/}}     

\title{Modeling Results and Baseline Design for an RF-SoC-Based Readout System for Microwave Kinetic Inductance Detectors}

\author[a,b,*]{Colm Bracken}
\author[b,c]{Eoin Baldwin}
\author[b]{Gerhard Ulbricht}
\author[b,c]{Mario de Lucia}
\author[b]{Tom Ray}
\affil[a]{Maynooth University, Department of Experimental Physics, Maynooth, Co Kildare, Ireland}
\affil[b]{Dublin Institute for Advanced Studies, Astronomy \& Astrophysics, 31 Fitzwilliam Place, Dublin 2, Ireland}
\affil[c]{Trinity College Dublin and CRANN, School of Physics, Dublin 2, Ireland}

\cftpagenumbersoff{figure}
\cftpagenumbersoff{table} 
\begin{document} 
\maketitle

\begin{abstract}
Building upon existing signal processing techniques and open-source software, this paper presents a baseline design for an RF System-on-Chip Frequency Division Multiplexed readout for a spatio-spectral focal plane instrument based on low temperature detectors. A trade-off analysis of different FPGA carrier boards is presented in an attempt to find an optimum next-generation solution for reading out larger arrays of Microwave Kinetic Inductance Detectors (MKIDs). The ZCU111 RF SoC FPGA board from Xilinx \cite{rfsoc1} was selected, and it is shown how this integrated system promises to increase the number of pixels that can be read out (per board) which enables a reduction in the readout cost per pixel, the mass and volume, and power consumption, all of which are important in making MKID instruments more feasible for both ground-based and space-based astrophysics. The on-chip logic capacity is shown to form a primary constraint on the number of MKIDs which can be read, channelised, and processed with this new system. As such, novel signal processing techniques are analysed, including Digitally Down Converted (DDC)-corrected sub-maximally decimated sampling, in an effort to reduce logic requirements without compromising signal to noise ratio. It is also shown how combining the ZCU111 board with a secondary FPGA board will allow all 8 ADCs and 8 DACs to be utilised, providing enough bandwidth to read up to 8,000 MKIDs per board-set, an eight-fold improvement over the state-of-the-art, and important in pursuing 100,000 pixel arrays. Finally, the feasibility of extending the operational frequency range of MKIDs to the 5 - 10 GHz regime (or possibly beyond) is investigated, and some benefits and consequences of doing so are presented.
\end{abstract}


\section{Introduction}
As part of the development of a new spatio-spectral focal plane instrument being designed, fabricated, and tested by a new low-temperature detector research group in Ireland, a new Frequency Division Multiplexed (FDM) readout electronics system was studied, employing an RF System-on-Chip (SoC). The MKID \cite{mkids1} fabrication work is done at Trinity College Dublin (TCD) and the Centre for Research on Adaptive Nanostructures and Nanodevices (CRANN). The cryogenic testing and characterisation is carried out at the Dublin Institute for Advanced Studies (DIAS), and the readout electronics research and numerical modelling is done by Maynooth University Experimental Physics and DIAS.

Low-temperature, superconducting microresonators are naturally suited for readout in the frequency domain. If a multi-gigahertz frequency range is utilised, this opens-up a large space for multiplexing huge numbers of extremely high-quality (high Q) resonators for a range of scientific applications. In contrast to, for example, Transition Edge Sensors (TESs) \cite{Irwin_2005} which have to combine the sensor with additional microresonators and Superconducting Quantum Interference Devices (SQUIDs) \cite{squid_res_1} to reach higher pixel numbers, multiplexing MKIDs is much more straight forward. It is difficult to put a number of the increase in pixel number possible with MKIDs compared to TES arrays. It would depend on many values, such as available volume in the cryostat, cooling power, and feedlines in/out of the system. What is more relevant is the reduction in complexity of reading MKID arrays compared to TES arrays. The removal of circuitry (for the SQUIDs) makes scaling-up to large pixel numbers far easier with MKIDs. In the case of kinetic inductance detectors, the microresonator itself often serves as the radiation sensor, where the inductor of the thin-film LC tank circuit acts as the sensitive area for the absorption of UV, optical, or even infrared radiation. By using an array of optical focusing elements such as microlenses \cite{len_kids_2_arcons,lens_kids_1,Meeker_2018}, the radiation can be focused onto the inductor portion of the resonator, effectively maximising the array fill-factor. MKIDs have also been successfully used to detect photons at far-infrared and submillimeter wavelengths, with alternative coupling schemes and quasioptical focusing elements such as feed horns \cite{horn_kids_1,horn_kids_2,Vissers_APL_2020}.

\begin{figure}[htbp]
\centerline{\includegraphics[width=90mm,height=70mm]{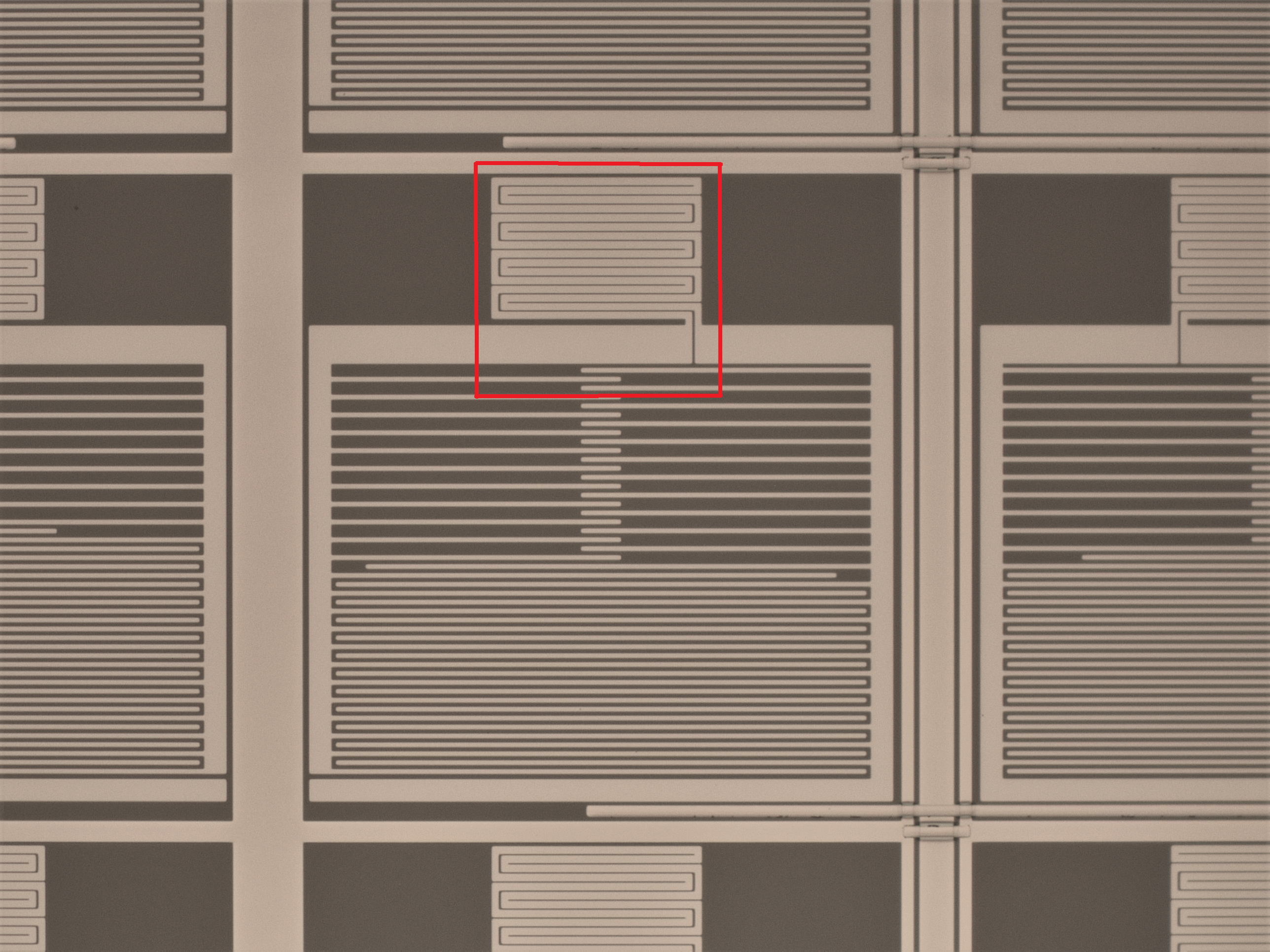}}
\caption{Close-up image of an MKID array, clearly showing the geometry of a typical lumped element MKID. The inductor (smaller meandering section marked with red boundary for clarity) and the larger inter-digitated capacitor (IDC) can be seen. It is the geometry of the capacitor that is typically varied from pixel to pixel, defining the unique resonant frequency for each pixel. Resonator design was based on work reported on in \cite{Szypryt_11}}.
\label{fig:lith2}
\end{figure}

Digital FDM readout electronics, based on Software Defined Radio (SDR) have already been shown to be a good solution for the readout of large-format arrays of optical/near-infrared MKIDs \cite{MAZIN_2006_2,mcHugh1,arcons1,Mazin12}. MKIDs, first demonstrated in 2003 \cite{mkids1}, are a specific class of cryogenic superconducting detector that show the most promise for scalabillity up to pixel number of $>100,000$ (Fig.~\ref{fig:lith2}). MKIDs are inherently multiplexable through FDM as each detector/resonator can be lithographically fabricated to have a unique resonant frequency in the GHz range. The basic principle for operating MKIDs requires cooling the microresonators below the critical temperature for superconductivity for their material ($T_c$), where the conduction electrons combine into so-called Cooper pairs \cite{Cooper_1956}. Typically we aim to maintain the resonators at a temperature of less than $1/8$ $T_c$. Thus, for a material with $T_c \approx 800$ mK, we cool the MKID device to 100 mK. Any photons with energy of more than twice the superconductor band gap incident on the material of a given resonator can cause breaking of the Cooper pairs, increasing the total inductance, which reduces the resonant frequency \cite{Zmuidzinas_2012}. An in-depth explanation of the kinetic inductance effect and superconductivity in general can be found in Zmuidzinas \cite{Zmuidzinas_2012} and Gao \cite{Gao_2008_2}, and some relevant aspects of the theory of superconductivity are covered by Mattis and Bardeen \cite{MattBard} and Kaplan et al. \cite{kaplan1}.

\begin{figure}[htbp]
\centerline{\includegraphics[width=90mm,height=70mm]{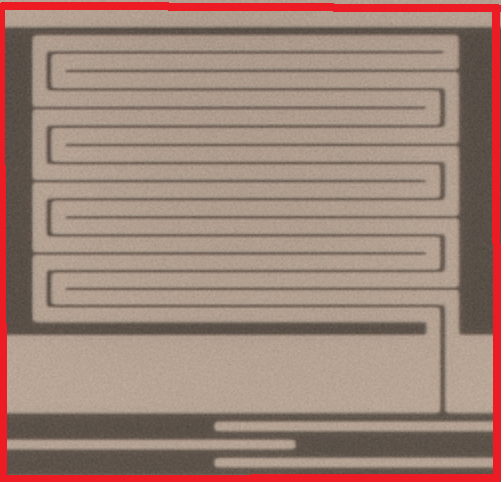}}
\caption{Magnification of the inductor section of a single resonator (marked-off by the red box in Fig.~\ref{fig:lith2}). This inductor gemoetry is typically kept constant for each resonator in an array, that each pixel should have the same sensitivity to the incident photons. Resonator design was based on work reported on in \cite{Szypryt_11}}
\label{fig:lith2_zoom}
\end{figure}

\section{Requirements and Methodology}\label{method}
\begin{figure}[htbp]
\centerline{\includegraphics[width=160mm,height=100mm]{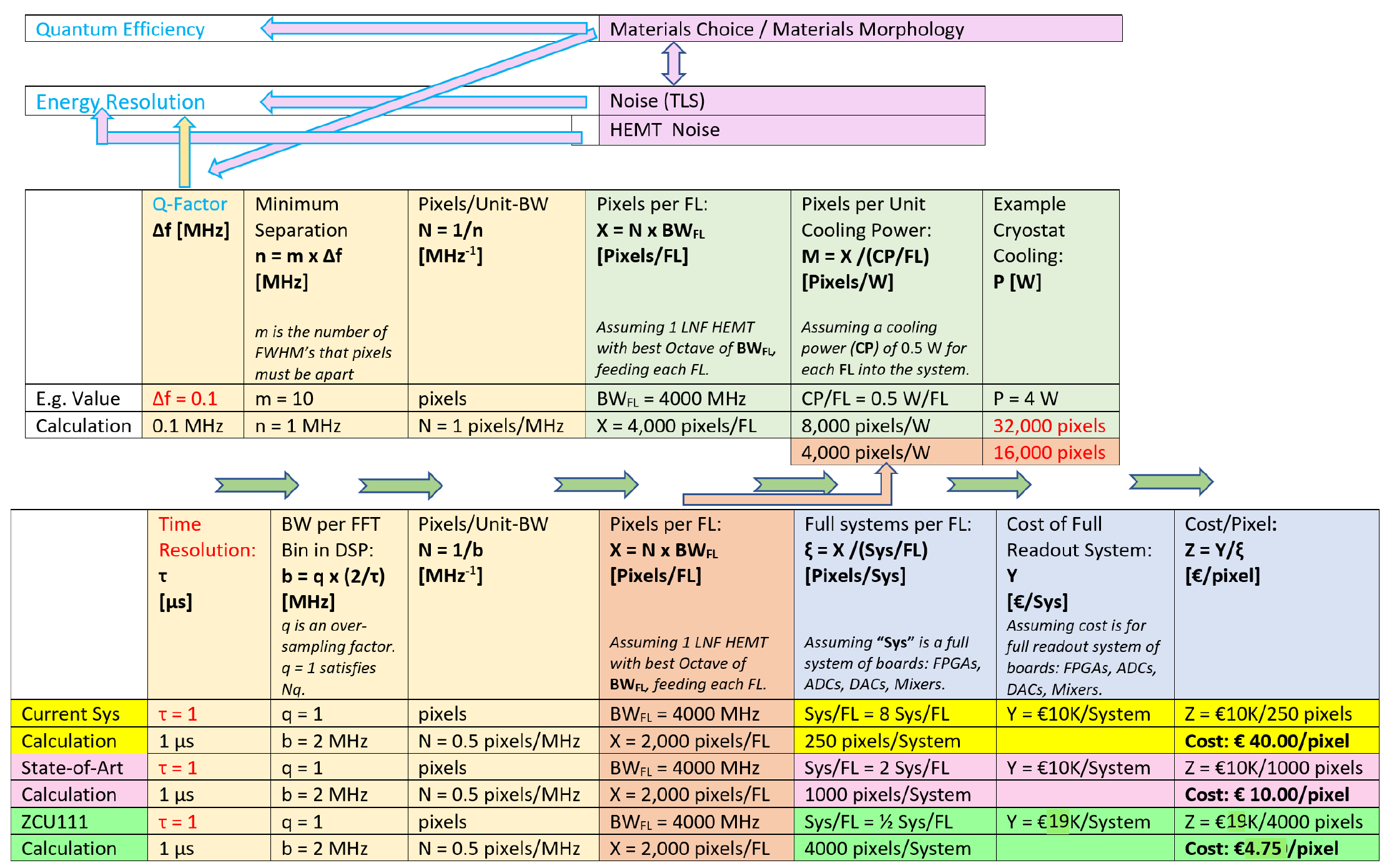}}
\caption{A diagram illustrating how the various MKID specifications, quality factors, and time-resolution requirements translate to readout specifications.}
\label{fig:complex_requirements_diagram}
\end{figure}

In this section we will layout the requirements we have set for our MKIDs, and how these requirements translate to readout specifications in terms of acceptable noise, channel isolation and flatness, sampling rates for the data converters, and bandwidth of the FFT bins. Fig ~\ref{fig:complex_requirements_diagram} illustrates the relatively complex manner of how some of these specifications and requirements of our MKIDs can define and constrain the specifications and cost-per-pixel of a readout system. 

\subsection{Time Resolution and ADC Sampling}\label{time_res}
An important point to state at the outset of the following sections, is that a relatively basic (and possibly naive) approach is taken in determining our acceptable noise levels. Based on the available literature to date for MKIDs designed for optical/near-IR photon detection,the assumption appears to be that the HEMT amplifier is the component in the signal line which sets the lower-limit for noise contribution. Based on this, we aim to achieve a noise floor equivalent to the measured noise temperature of our HEMT, namely $\approx$2.5 K.

\begin{figure}[htbp]
\centerline{\includegraphics[width=90mm,height=55mm]{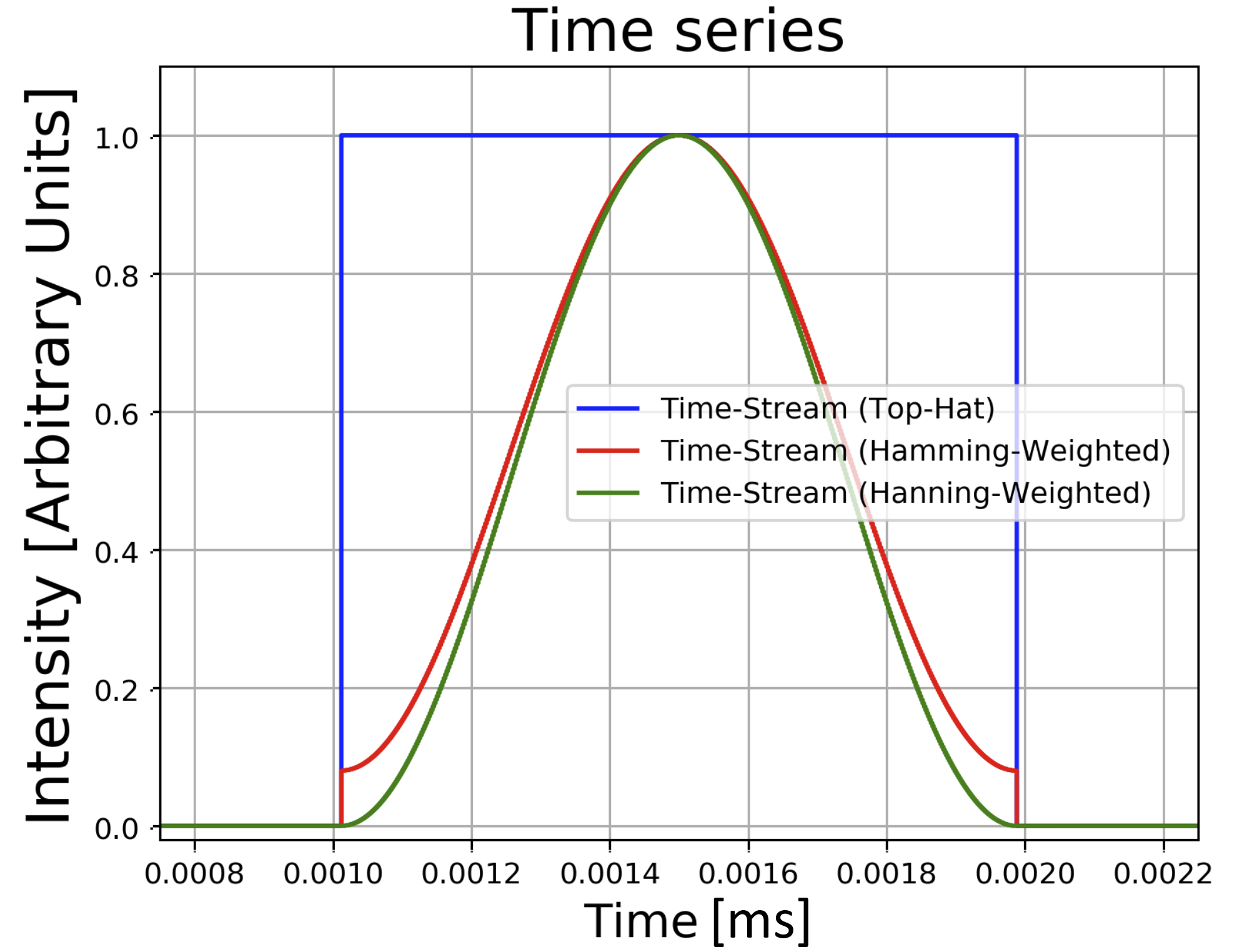}}
\caption{Simulated time-stream data for a raw data stream truncated with a standard top-hat, as well as weighted data corresponding to two different filters. In each case the sampling rate was 4 GSPS, with 1 microsecond of non-zero values defined, and padded by zeroes either side.}
\label{fig:time_streams_1}
\end{figure}

The particular frequency regime targeted with any MKID-based instrument will play an important role in determining many of the DSP parameters. Since we are developing a relatively high-format array of MKIDs intended for photon-counting in the optical to near-infrared range, and we desire an ability to time-stamp each photon to a precision of roughly 1 microsecond, our channelisation firmware will have quite strict requirements.
The microsecond time resolution obviously requires that we effectively sample each pixel one million times per second. Since we intend to implement an FDM technique, which involves buffering a series of time-sampled data before applying an FFT, we must ensure that each FFT `loop' repeats with a recurrence frequency of $\approx 1 \mu$s. A simulated time-stream data `chunk' is shown in Fig.~\ref{fig:time_streams_1}.

\begin{figure}[htbp]
\centerline{\includegraphics[width=90mm,height=55mm]{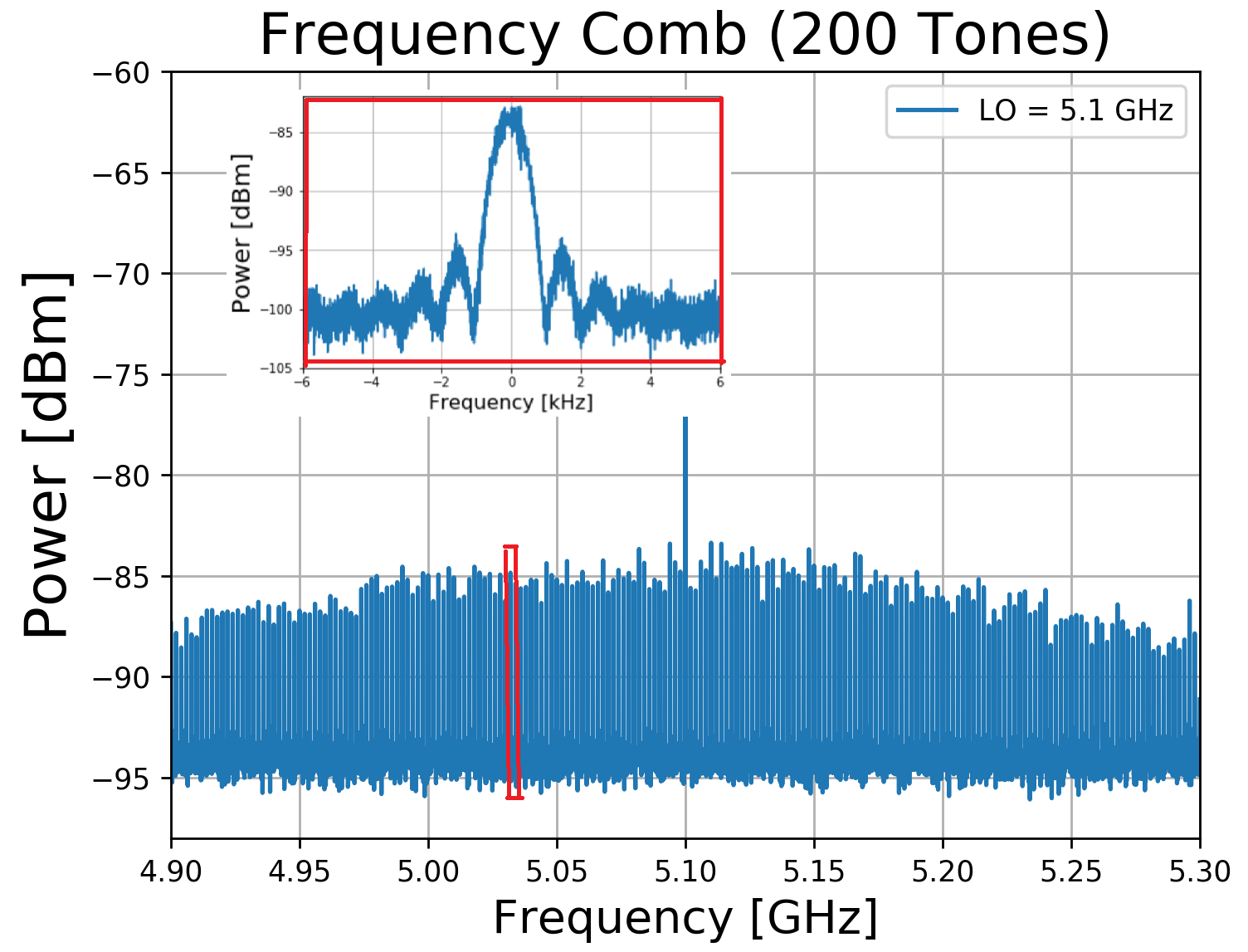}}
\caption{A typical broadband frequency comb waveform, measured with a VNA.}
\label{fig:comb_vna}
\end{figure}

Based on the required time-resolution, for each GHz of bandwidth from a single ADC (or, 1 Giga-Samples Per Second (GSPS) available), we can expect to achieve a number of frequency bins ($N_B$) of roughly:
\begin{equation}
N_B = \frac{1\times 10^{9} s^{-1}}{2\times 10^{6} s^{-1}}\label{eq:sampling_1} \approx 500
\end{equation}
where the factor 2 in the denominator is to satisfy the Nyquist criterion.
Each of these 500 frequency bins will be sensitive to 1 MHz of bandwidth, ranging from [0-1, 1-2, ..., 498-499, 499-500] MHz.
However, since we will synthesise our output tones using two DACs (one for In-phase ($I$) waveforms and one for Quadrature ($Q$) waveforms), and we will use two ADCs for digitising the complex signals before processing, we can also define negative frequency baseband probe signals and FFT bins, respectively. These negative frequency bins will range from -500 MHz to 0 MHz. The positive DAC baseband tones (rotating counter-clockwise in the I/Q plane), when up-mixed to higher frequencies with an LO and I/Q mixer, will yield the RF tones for driving the MKIDs in the upper side-band of up-mixed output. The negative baseband tones (rotating clockwise in the I/Q plane) will yield the RF tones in the lower side-band of up-mixed output. A plot of an example broadband frequency comb comprising a few hundred tones uniformly separated by 1 MHz, after up-mixing with an LO of 5.1 GHz, is shown in Fig.~\ref{fig:comb_vna}. Correspondingly, after the RF tones exit the MKIDs array, and are down-mixed with the same LO frequency, they can be channelised in both positive and negative FFT bins. Considering the above constraints, a pair of 1 GSPS ADCs will allow just over 1,000 frequency bins, each sensitive to 1 MHz bandwidth.

\subsection{Sensitivity and Pixel Cross-Talk}\label{bin_resp}
\begin{figure}[htbp]
\centerline{\includegraphics[width=90mm,height=55mm]{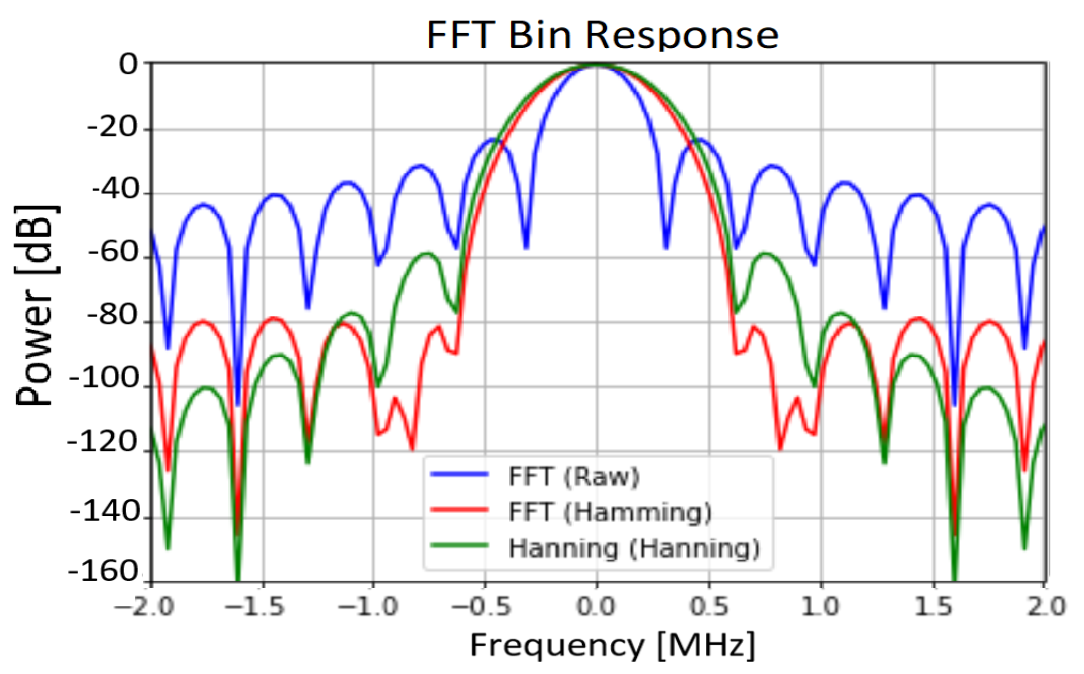}}
\caption{Power attributed by a discrete FFT of a perfectly sharp input signal to a single, 1 MHz wide frequency bin, plotted over the deviation of input signal frequency to bin center, using the different time-stream data shown in Fig.~\ref{fig:time_streams_1}.}
\label{fig:freq_streams_1}
\end{figure}

While a raw FFT could be employed for some FDM applications, there are a number of potential issues that arise when employing the FDM readout technique for MKIDs, particularly if we wish to pack the MKIDs close together in frequency space. So-called clashing pixels, or pixel cross-talk, where two or more pixels are too close in frequency, have been a primary cause of low pixel yield in previous optical/near-IR MKID instruments such as ARCONS and DARKNESS \cite{arcons1, Meeker_2018}. The clashing typically results when the measured resonant frequencies shift significantly from the design frequencies. These resonance shifts can be due to minor fabrication inaccuracies, or resonators which exhibit a drift in resonant frequency over time. Ideally, we would like to have each pixel's resonant frequency to be precisely centred on one of the frequency-domain bins, with a 2 MHz spacing between neighbouring pixels, but this is difficult to achieve. It would also be desirable for each resonator to be optimally driven with an equal power level, but again, this is not typically observed. Considering these non-ideal frequency-spacing and driving-powers, a raw FFT would appear to be offer insufficient isolation from pixel cross-talk or clashing. A raw FFT will typically exhibit sidelobe structure with sensitivities approaching -20 dB (relative to bin-centre sensitivity), as shown in Fig.~\ref{fig:freq_streams_1}. This means that if a pixel happens to resonate at a frequency corresponding to the side-lobe of a neighbouring pixel, any significant change in signal throughput (say, from an incident photon causing the pixel to shift off-resonance) may appear as signal increase in the neighbouring pixel too. Now, we aim to drive our resonators with relative power of up to 20 dB, where relative power refers to the change in power throughput of a probe tone when a resonator is switched from being driven on-resonance to off-resonance. Further, the actual power required to optimally drive each resonator, typically around -100 dBm, can vary from pixel to pixel by up to $\pm$10 dB. The most obvious way to reduce the levels of these cross-talk effects is to drive down the side-lobe structure in the FFT bins. This is best achieved by implementing a digital window function. The significant difference this can make is illustrated in Fig.~\ref{fig:freq_streams_1}. For robustness, we aim to ensure that the highest side-lobe in our FFT bins is no larger than -40 dB (relative to bin-centre sensitivity). An additional benefit of this side-lobe suppression is to reduce our sensitivity to higher frequency noise. This is also very important for our applications, since we aim to maximise energy resolution, $\Delta E$, and this is directly affected by signal to noise ratio (SNR).

\subsection{Signal-to-Noise Requirements, and Scalloping Losses}\label{SNR_req}
Another consideration given to the structure of our FFT bins is flatness of response of each bin. Given that we aim to drive our resonators at relatively low powers of around -100 dBm, and the fact that SNR at these low power levels is challenging, we want to ensure that we are fully sensitive to the signal power of each probe tone. Since we can not expect that each of our resonators will always fall precisely on a bin centre, as per design, the non-flat response of the FFT bins, even with an appropriate window function applied, is a concern. We thus desire a perfectly flat response for each FFT bin. Another issue, related to the non-flat bin response, is so-called scalloping losses. Small gaps between FFT bins are typically present if additional DSP steps are not implemented. We correct for these potential scalloping losses and non-flat bin respnose by employing a polyphase filterbank (PFB). Unfortunately there is a correlation between the effectiveness of the PFB and the amount of FPGA resources required. More importantly, there is a direct correlation between the effectiveness of the PFB and the time resolution in the final data. These questions and trade-offs are discussed now.

\begin{figure}[htbp]
\centerline{\includegraphics[width=90mm,height=55mm]{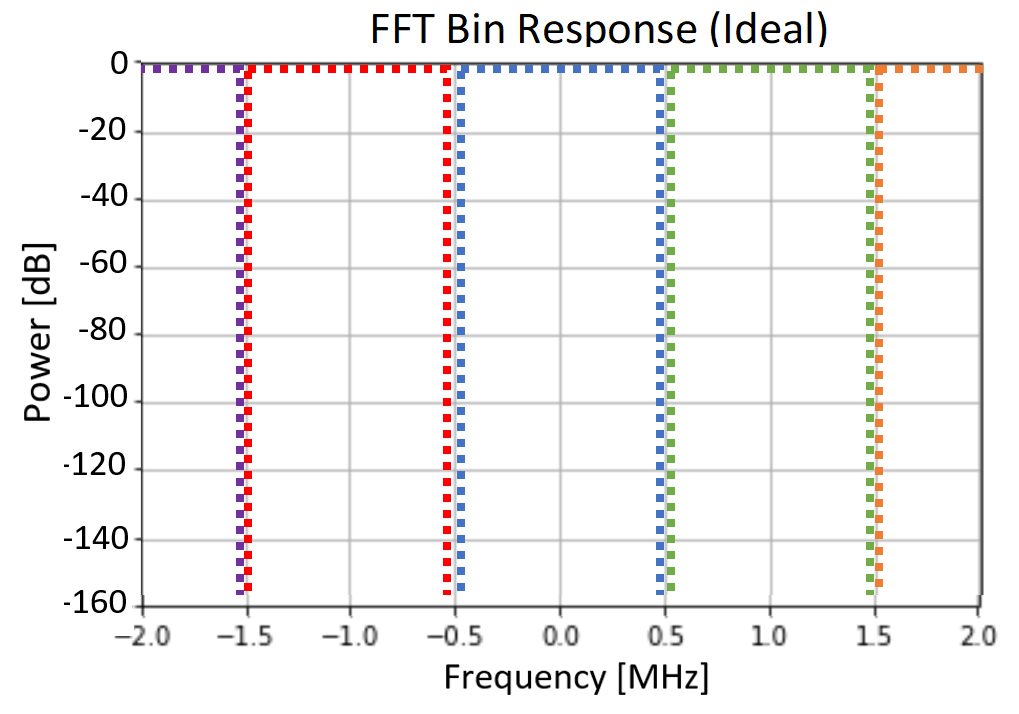}}
\caption{The ideal frequency response of our DSP, with perfect isolation and no truncation of signals.}
\label{fig:tophat_ideal}
\end{figure}

There are a number of techniques we could apply to approximate the perfectly flat, perfectly isolated bin response shown in Fig.~\ref{fig:tophat_ideal}.
One approach would be to use a large number of taps in the PFB. The FFT of a top-hat function is of course the function. Therefore, we should be weighting or `windowing' each group of time-stream data points $g(t_n)$ by coefficients $(h_m)$ corresponding to a digital filter, essentially applying a sinc-type normailisation. As such, in the frequency domain we would like to see the effect of $\mathcal{F}\{g(t_n)\}\cdot\mathcal{F}\{h_m\}$. However, the convolution theorem tells us that if we want to apply this in the time-domain, the problem becomes a discrete convolution as shown in Eq.~\eqref{eq:FFT_Conv}.
\begin{equation}
\mathcal{F}\{g(t_n)\}\cdot\mathcal{F}\{h_m\} = \mathcal{F}\{g(t_n)\circledast h_m\} \label{eq:FFT_Conv}
\end{equation}

\begin{figure}[htbp]
\centerline{\includegraphics[width=90mm,height=55mm]{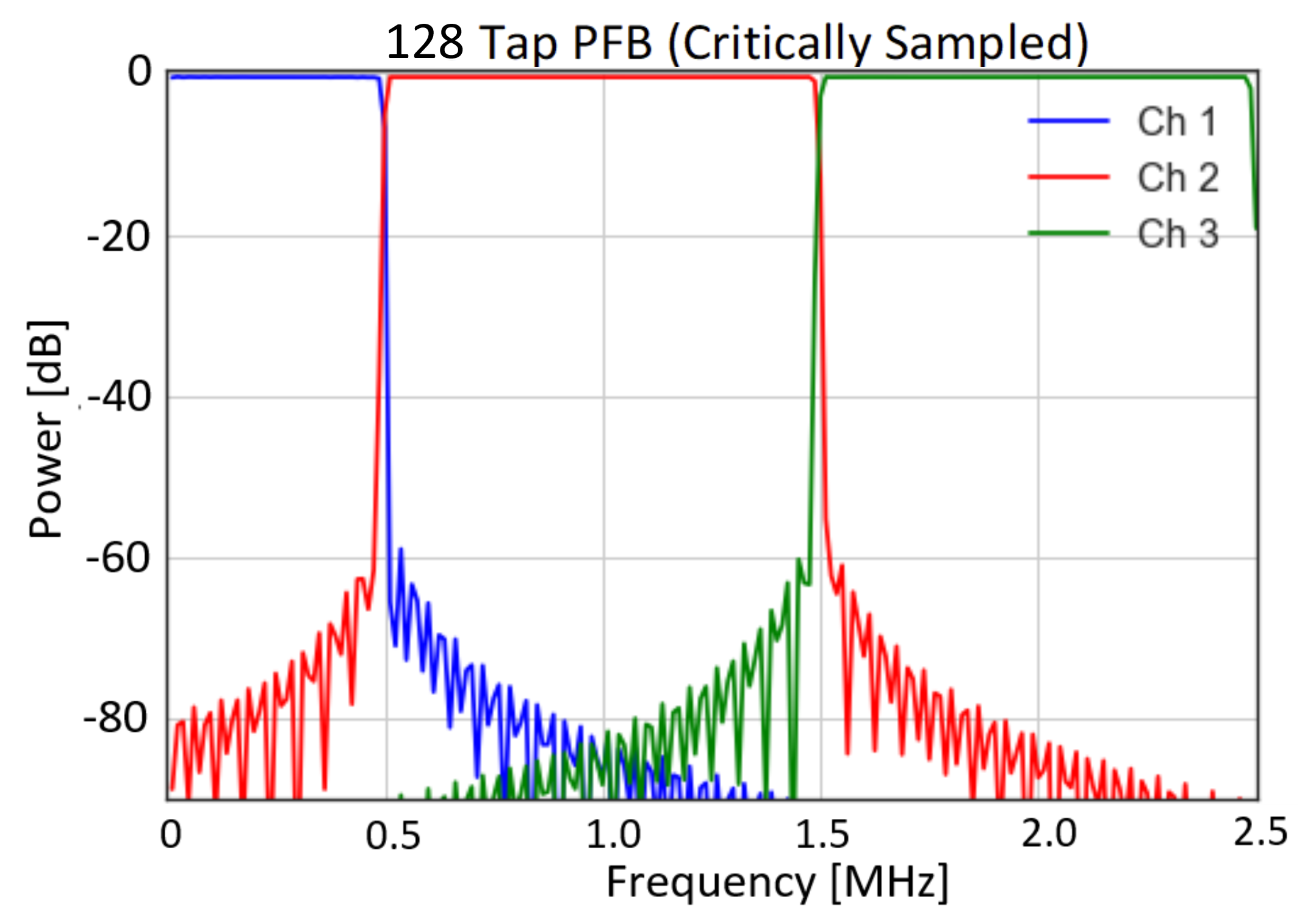}}
\caption{Simulated response of a 128-tap PFB (with a Hamming FIR window before the FFT). Three adjacent channels are shown, indicating the expected levels of spectral leakage, and losses due to scalloping.}
\label{fig:128_tap_pfb_hamming}
\end{figure}

Looking at he right hand side of Eq.~\eqref{eq:FFT_Conv}, it is clearly a discrete convolution if $n = m$. The authors have reviewed the literature, and have not found it explained as such before. This is most likely due to the fact that in practice, $m$ is usually set to a value much smaller than $n$, and this would then describe a traditional PFB. For the system requirements described above, this would be an $n = m = 1024$ for each 1 GSPS of sampling. This would require buffering 1024 blocks of data time-streams - each 1 $\mu$s in length and containing $\approx$1000 samples - and weighting each block with its corresponding coefficient $h_m$ of the FIR window function. The weighted blocks would then be summed, and the result would undergo the FFT. Now, the implication most discussed in the literature of such PFB systems is the increased amount of logic and memory for buffering, which we are indeed trying to minimise on our chip. Nonetheless, a moderately large PFB was simulated with $m = 128$, and the result is shown in Fig.~\ref{fig:128_tap_pfb_hamming}. As can be seen the FFT bins are already quite uniform and isolated even with $m = 128$.

A more interesting implication to using a large number of PFB taps is the effect that this has on the final time resolution. The buffering and summing of the large number of data-streams essentially washes out the time information which a more simple windowed-FFT would yield. And, given that we are aiming for roughly microsecond time resolution, the number of taps we apply in the processing should be considered carefully. The indirect effects on MKID energy resolution will need to be examined also. This work will form the basis of a later paper.

\begin{figure}[htbp]
\centerline{\includegraphics[width=90mm,height=55mm]{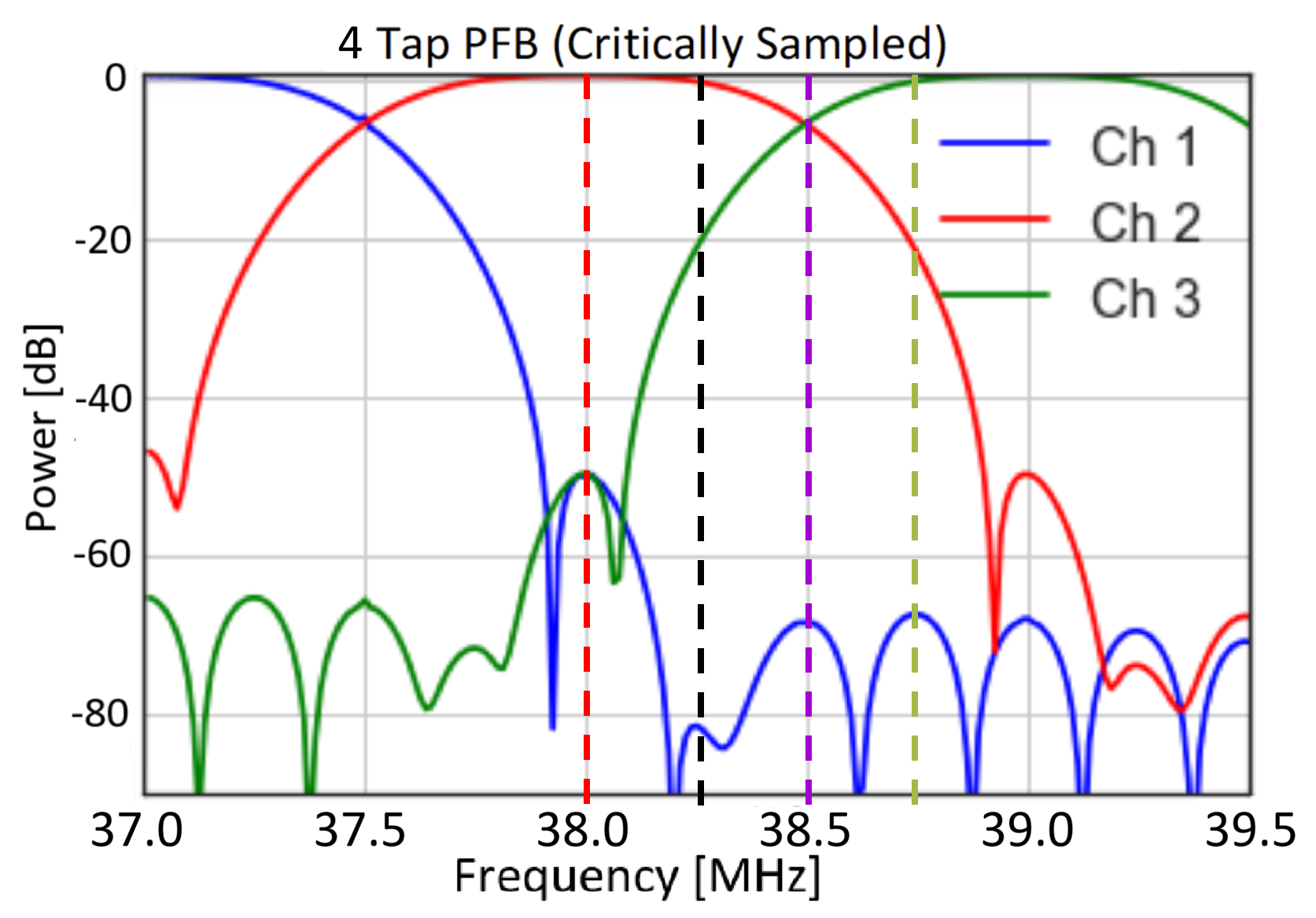}}
\caption{Frequency response of a 4-tap PFB which employs a Hamming FIR window before the FFT. Three adjacent channels are shown, indicating the expected levels of spectral leakage and losses due to scalloping. The 4 dashed vertical lines indicate the frequencies of 4 test tones that were measured, as shown in Fig.~\ref{fig:pfb_scallop_meas_sim}.}
\label{fig:pfb_hamming}
\end{figure}

For now, we settle for a much lower number of taps, and only crudely approximate the top-hat shape. Fig.~\ref{fig:pfb_hamming} shows the frequency bin response of a 4 tap PFB with the commonly used Hamming FIR window. For the reader unfamiliar with these DSP techniques, there are a number of sources available in the literature; \cite{danny_price_1}, for example.

\begin{figure}[htbp]
\centerline{\includegraphics[width=90mm,height=55mm]{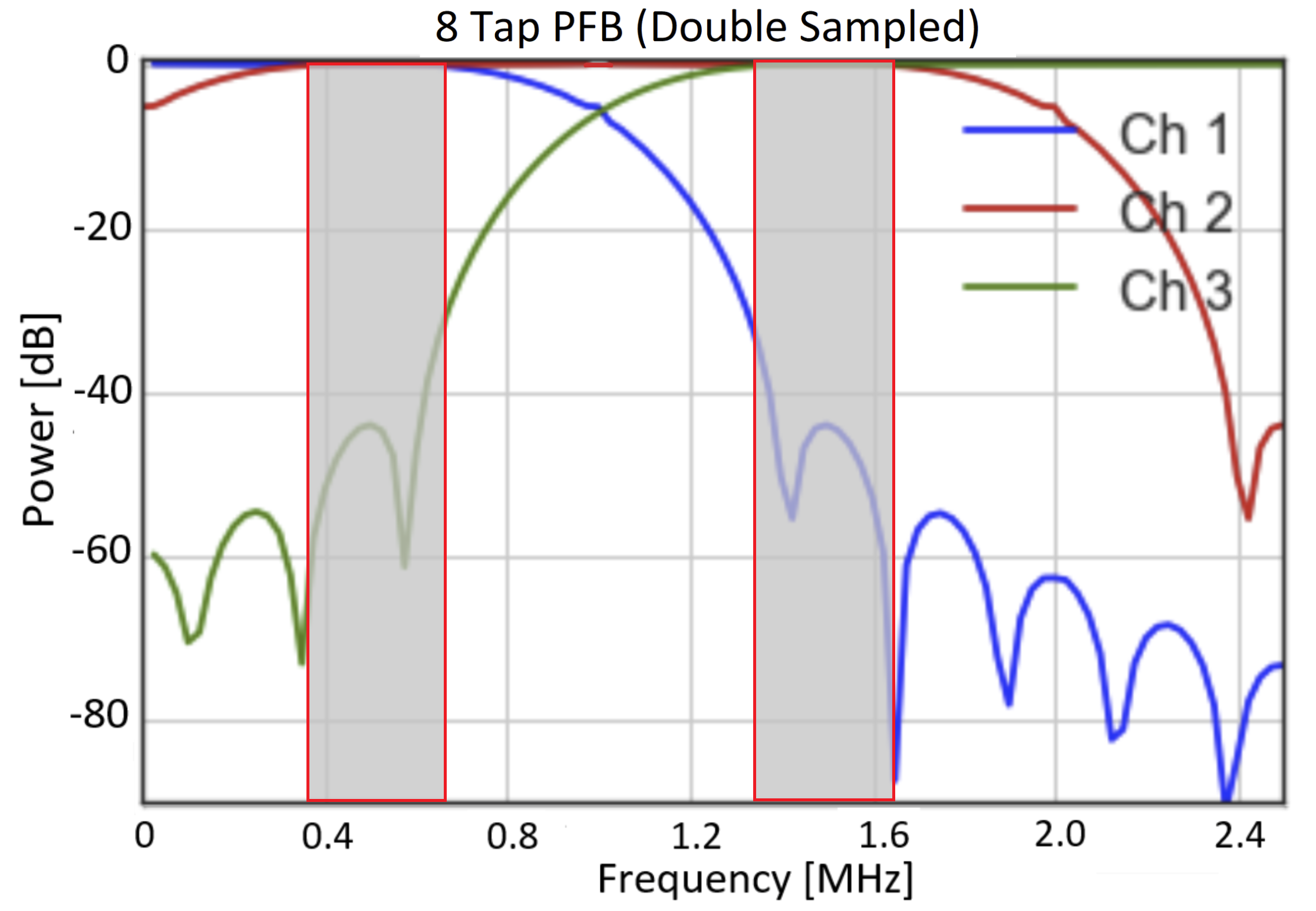}}
\caption{Frequency response of an 8-tap PFB which employs a x2 window length, and would require two staggered FFT stages to double up the sampling if aliasing is to be avoided. A Hamming FIR window function was used.}
\label{fig:double_samp_1}
\end{figure}

\begin{figure}[htbp]
\centerline{\includegraphics[width=90mm,height=55mm]{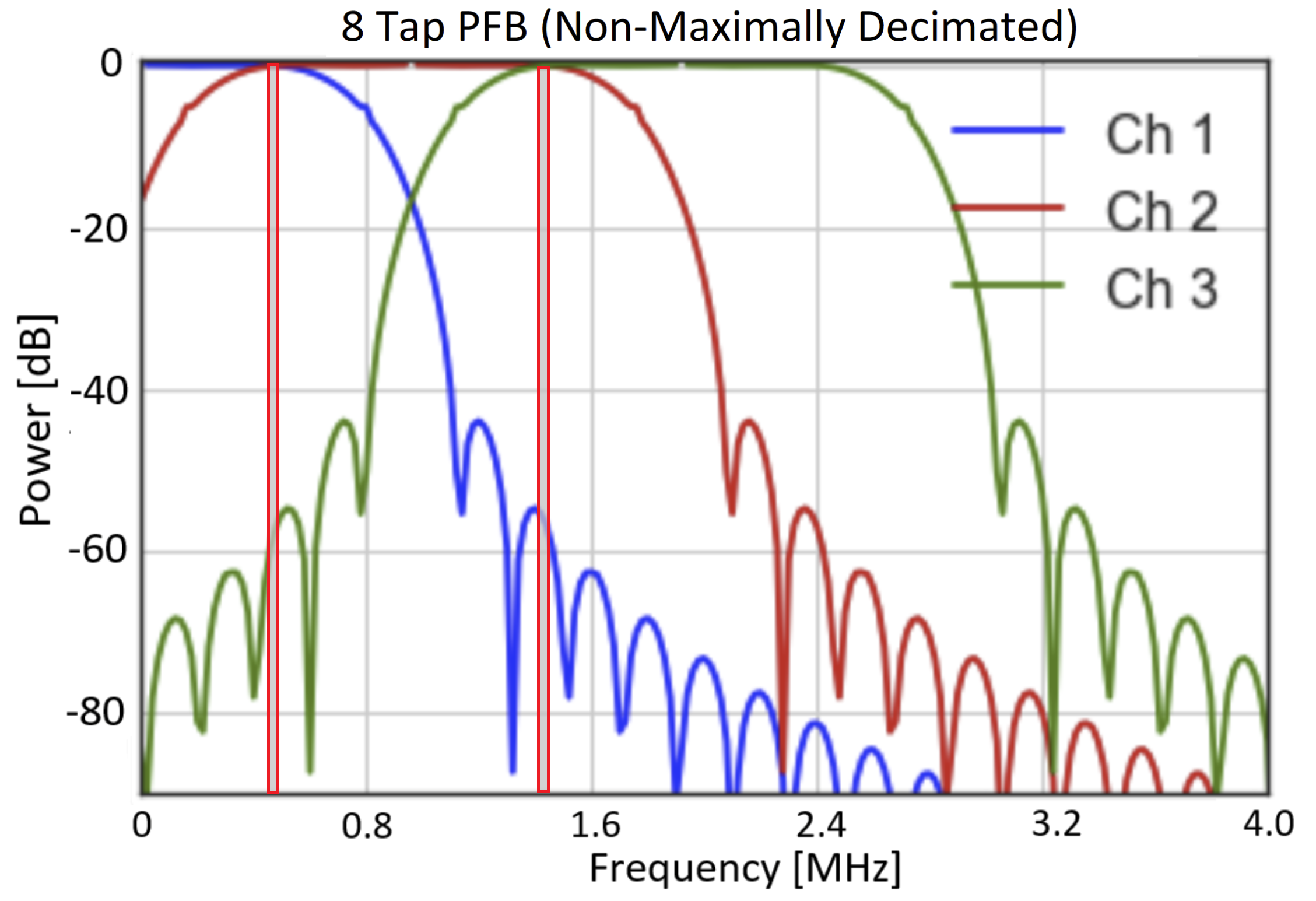}}
\caption{Frequency response of an 8-tap PFB which employs a slightly increased window length (32/27 x critical). This approach only requires partial oversampling (or non-maximally decimated sampling). A Hamming FIR window function was again used.}
\label{fig:32_samp_1}
\end{figure}

The scalloping losses due to the sensitivity dips between neighbouring FFT bins are a main concern for our application, as the signals driving the MKID resonators are typically very low power, as previously mentioned. The scalloping losses could be taken care of in a number of ways, including:
\begin{enumerate}
    \item{Simply use a standard, critically Nyquist-sampled PFB, and calibrate/normalise for the lower signal levels since the level of scalloping loss will be determinable.}
    \item{Implement overlapping FFT bins, covering-over the dips between FFT bins, and thereby removing scalloping loses. This is achieved by doubling the FIR window function width, and then using two overlapping (but out of phase) FFTs, effectively increasing the sampling rate back up to satisfying Nyquist criteria.}
    \item{Implement a partially oversampled (or so-called `non-maximally decimated' PFB, with a partially widened FIR window function.}
\end{enumerate}
Of the three solutions above, option 1) is the easiest to implement, and the most computationally efficient. However, the signal-to-noise ratio will still be affected, and this is unacceptable for our application. Option 2) is also relatively straightforward to implement, and has been the standard approach used for optical/near-IR MKID readouts \cite{mcHugh1,strader1}. However, it is very demanding in terms of FPGA logic. Fig.~\ref{fig:double_samp_1} shows an implementation of option 2) for three adjacent channels, where the overlapped channels are now 2 MHz wide but still separated by 1 MHz, yielding a 50\% overlap. However, as shown by the shaded grey areas there is quite significant full uniform overlap where there is uniform response across roughly 0.3 MHz of bandwidth covered twice in each channel. This area of uniform overlap slightly increases as the number of taps of the PFB is increased, and it is deemed overkill and wasteful of FPGA resources. As such, an alternative, more economic approach was sought. Option 3) is the most difficult technique to implement in firmware, but it requires less logic resources than option 2, while still providing a flat response across the full band. It has been found by J. Tuthill, et al., 2015 \cite{OverSampCorr_2017}, for example that a non-maximally decimated (or partially oversampled) PFB is feasible to implement. If a  time-stream longer than the critically sampled points is fed into the FFT in combination with a widened FIR window function, this widens the FFT bin response by approximately 20\% and can sufficiently `fill the gaps' between FFT bins, ensuring no scalloping losses occur \cite{OverSamp_2012,OverSampCorr_2017}. Then, an oversampling ratio of 32/27 satisfies the Nyquist criteria for the now broader frequency response of the FFT bins. Fig.~\ref{fig:32_samp_1} shows a simulation of the 32/27 non-maximally decimated PFB, with minimal redundancy in uniformity of response shown by the very narrow shaded regions.

Using the non-maximally decimated approach should save significant logic resources on the FPGA. However, while this non-maximally decimated PFB does indeed flatten the full passband, it has also been identified that this partial oversampling introduces so called phase rotations \cite{OverSampCorr_2017}: If a signal analyzed with a discrete FFT with a finite data sampling rate is not at an integer multiple of that sampling rate, the resulting power in a FFT bin will oscillate with the difference between the bin center frequency and the signal frequency. This oscillation is removed in the last DSP step, the low pass filtering (see below). A non-maximally decimated PFB slightly shifts the FFT bin centers, thereby introducing a periodic phase rotation for any 'would-be' bin-centred signal. The velocity (or frequency) of this rotation will be dependent on the particular FFT bin and where it lies within the full band. The phase-velocity will equal the difference between the FFT bin centre and the shift introduced by the non-maximal decimation Eq.~\eqref{eq:res_shift_1}. A detailed explanation of the cause of this phase-rotation, and a solution to correct for it is discussed in section \ref{fine_chan}.
\begin{equation}
\phi_f = f_0 - f_{shift}\label{eq:res_shift_1}
\end{equation}

\subsection{Basic Experimental Verification of Models}\label{SNR_req}
A primary objective of this work was how we might optimise the efficiency of the DSP; that we might take full advantage of available FPGA resources, and thus maximise the number of MKIDs to be read per board set. However, editing firmware in the Xilinx Vivado tools, then compiling the software into bit-stream files, and finally loading the bit-streams onto the FPGAs is very time consuming. For this reason a Python DSP simulator was developed, so that new DSP designs and ideas could be quickly iterated through, purely in software. While most of this simulation work was based on this in-house DSP simulator, additional open-source code was also used for comparison of our PFB results \cite{danny_price_1}.

\begin{figure}[htbp]
\centerline{\includegraphics[width=90mm,height=55mm]{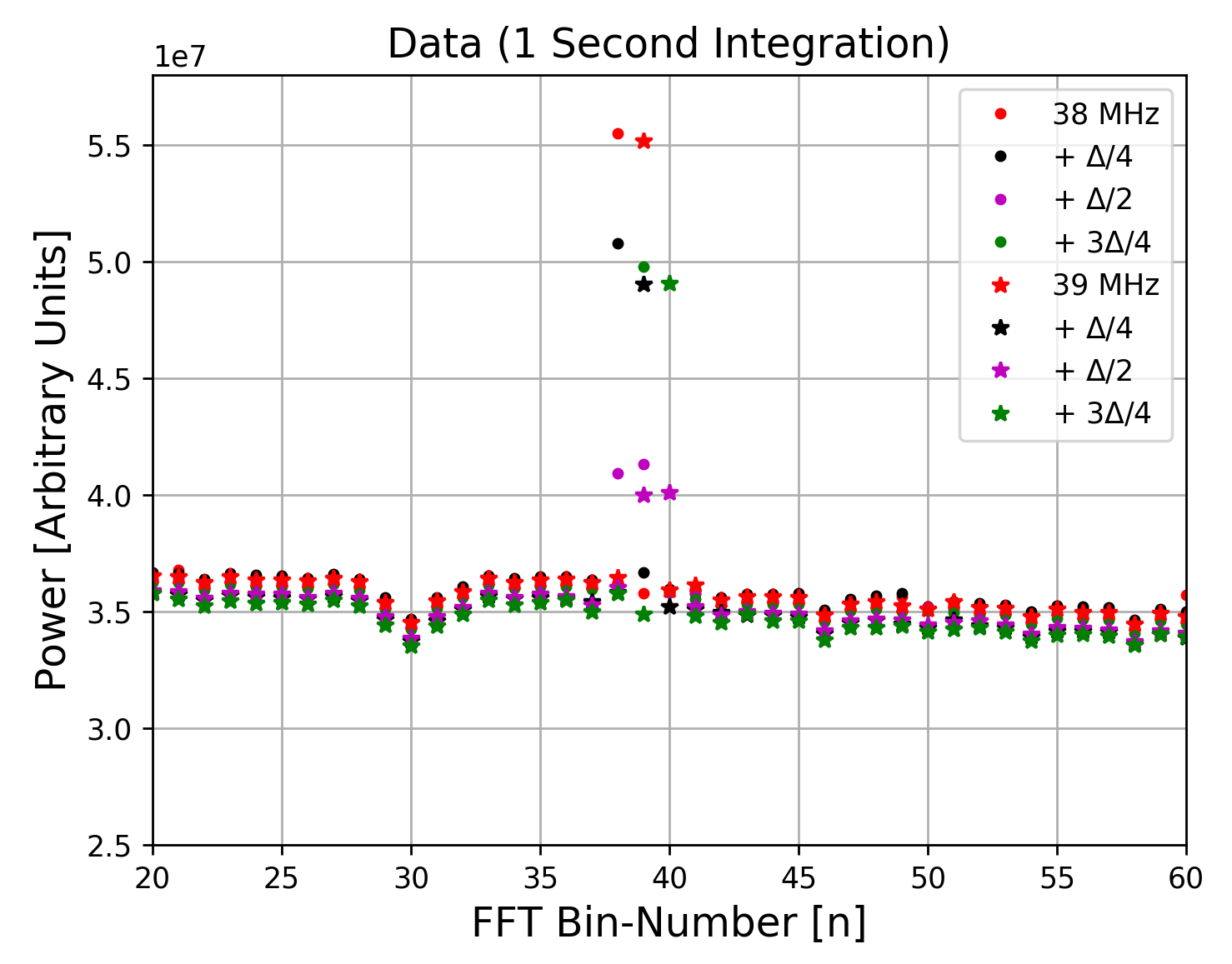}}
\caption{A number of narrow-band tones were channelised with a 4 Tap PFB. Tones around 38 MHz and 39 MHz were used, with increasing in frequency in steps of $\Delta/4 = 0.25$ MHz. A 1 second integration is shown here, ten of which were used and averaged.}
\label{fig:pfb_scallop_meas}
\end{figure}

\begin{figure}[htbp]
\centerline{\includegraphics[width=90mm,height=55mm]{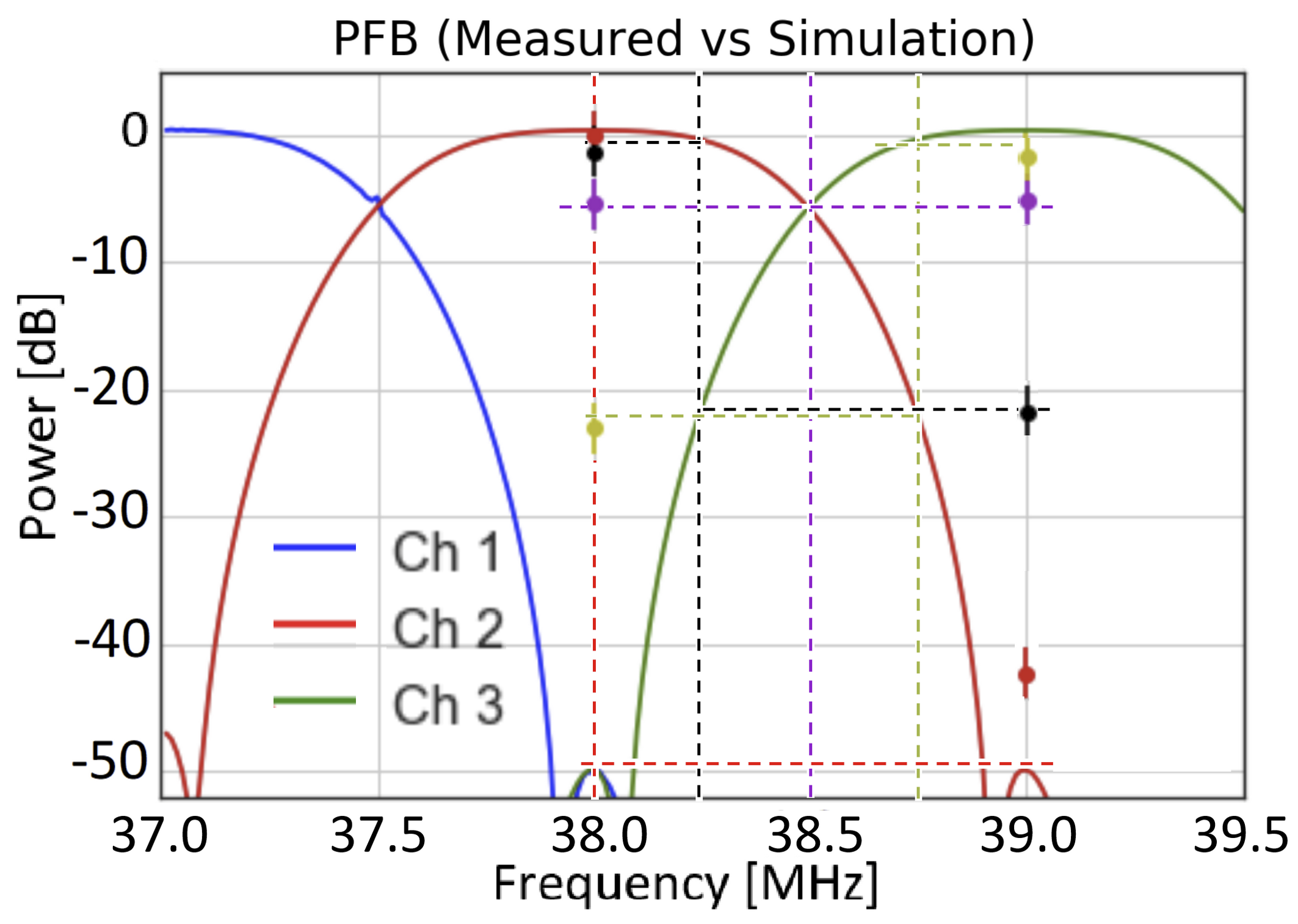}}
\caption{Zoomed-in data from Fig.~\ref{fig:pfb_scallop_meas}, after calibration and normalisation.}
\label{fig:pfb_scallop_meas_sim}
\end{figure}

An experimental verification was carried out for the predicted losses due to scalloping effects on non-bin-centred tones. The measurements also served to check the inter-channel leakage predicted by the models. The FPGA was programmed with a critically sampled, 4 tap, 256 frequency-bin PFB (using a Hamming FIR window). Using a signal generator, a series of tones were then fed to the ADC, where the tones were given frequencies to be centred on two FFT bins, and then stepped-up in increments of $\Delta/4$, where $\Delta = 1$ MHz (the frequency spacing between bins). The system was stable and reproducible to within $\approx\pm$1.1 dB, which was considered sufficiently stable (relative to the somewhat approximate requirements laid out in sections \ref{bin_resp} and \ref{SNR_req}). Fig.~\ref{fig:pfb_scallop_meas} shows the raw data from the measurements. Although the raw data shown has not yet been normalised and corrected, the effects of the roughly -6 dB edge-of-frequency-bin scalloping predicted by the model (Fig.~\ref{fig:pfb_hamming}) can be seen to clearly reduce the signal for tones not centred on a given bin. Fig ~\ref{fig:pfb_scallop_meas_sim} shows a more detailed analysis of the measurements, with the PFB bins overlaid for comparison.

Clearly in Fig ~\ref{fig:pfb_scallop_meas_sim} the bin-centred tone (red) is picked up perfectly by the 38 MHz bin, and partially detected by the bin to the right. The tone with 38.25 MHz ($f_0+\Delta/4$) (black) is very slightly attenuated due to the non-uniform response across the sub-channel, with the neighbouring bin again picking up the signal (a bit stronger this time). Then, with the 38.5 MHz ($f_0+\Delta/2$) tone (purple), it is attenuated by $\approx 6$ dB, and is detected at identical levels in the two adjacent bins, since it falls exactly between the two bins. Finally, for the 38.75 MHz ($f_0+3\Delta/4$) tone (yellow), we have the opposite case as for the 38.25 MHz tone, namely a slightly attenuated signal in the 39 MHz bin, and a heavily attenuated signal in the 38 MHz bin. This makes sense since this is equivalent to 39 MHz - $\Delta/4$.

There is nothing too surprising about these results, but the good agreement between measurement and simulation simply verifies that our Python DSP simulator is working correctly. The verification of our models was pleasing, as it provided confidence to push forward with new DSP designs. Writing DSP firmware, then transferring it to FPGAs and other hardware, and then testing it is very time consuming. Instead, simulating new prototype DSP designs in software is far faster, but of course it is important to verify these simulations regularly.

The only measured data point that is significantly higher than the simulated value is the inter-channel leakage of the bin-centred tone (red) into the neighbouring 39 MHz bin. This indicates that the first side-lobe of the channels may be a bit higher than that predicted by the model. In fact, through these simulations it was noticed that when a 4 tap PFB is implemented, the first side-lobe of each sub-channel falls directly on the centre of the neighbouring sub-channel, regardless of the FIR window function that is used (Hamming, Hann, Blackmann-Harris, etc). Nonetheless, the leakage was still recorded to be below the required -40 dB level, so considered to be acceptable. Furthermore, the measured spectral leakage between frequency bins is not a significant concern, as it will be removed automatically by subsequent steps in the DSP pipeline (lowpass filtering after digital down conversion).

\section{Trade-off Comparison of Different Hardware Solutions}\label{approach}
Due to the recent successes of Field Programmable Gate Array (FPGA) carrier boards being used for optical/near-IR MKID readout \cite{mcHugh1,strader1}, we decided to look for a similar (but improved) solution for our next-generation board set. A board set, as defined here, comprises analogue-to-digital and digital-to-analogue data converters (ADCs and DACs), processing units (FPGAs, CPUs or GPUs), and IF/RF mixers used for up/down frequency conversion. The primary aims for a next-generation board set are:

\begin{enumerate}
    \item{An increase in the number of pixels that can be readout and processed (per board set).}
    \item{A reduction in mass and volume of a full board set.}
    \item{A reduction in the cost of a full board set (which then translates to a reduction in the cost per pixel, based on point (1) above).}
    \item{A reduction in power consumption of each board set.}
\end{enumerate}

The successfully commissioned ARCONS MKID instrument \cite{len_kids_2_arcons}, and more recently the DARKNESS MKID instrument \cite{Meeker_2018} for optical/near-IR wavelengths used the ROACH 1 and ROACH 2 boards \cite{casper_10_years}, respectively, as the core of their readout systems. These ROACH boards had originally been designed for radio astronomy by the Collaboration for Astronomy Signal Processing and Electronics Research (CASPER) group \cite{casper_cite_1}. As such, the most obvious boards to examine first were the current cutting-edge bespoke FPGA carrier boards being used by the radio astronomy community. The SKARAB board \cite{peralex_1} by Peralex (often nick-named the ROACH 3) is currently the most powerful CASPER-supported FPGA system in use in radio astronomy. Now, although the SKARAB offers impressive processing capacity in terms of FPGA logic, there were a number of issues identified in its application as an MKID readout system. First, there are no DAC units available to fit the customised Mezzanine port, so an adapter unit would be required to convert commercially available DACs. Currently, the most commonly used connection for high-throughput data transfer for industry standard data converters is the FPGA Mezzanine Card (FMC). Adapting high-speed FMC-based DACs with the SKARAB's Mezzanine port would require non-trivial hardware development with additional costs and potential compatibility complexities. Furthermore, the SKARAB and its custom ADCs were very expensive, resulting in a very high cost per pixel which ultimately directed us away from this option.

Other bespoke systems designed with radio astronomy in mind were examined, and compared, but all required at least some level of hardware alteration, thereby driving up the readout cost per pixel. This should have been expected, as even the ROACH 2-based DARKNESS readout required additional FPGAs and hardware alterations. Having exhausted the bespoke electronics solutions available, we next looked to `commercial off-the-shelf' (COTS) options. The readout electronics for MKID instruments continually benefit from the regular advances in high speed data converters and signal processing capabilities being demanded by industry; 5G technology being the latest and most demanding example. To satisfy some of this demand, in early 2019 Xilinx Inc. released their range of System-on-Chip (SoC) FPGA series, including the Zynq UltraScale+ RF-SoC \cite{rfsoc2}. These chips, which contain generous logic and Digital Signal Processing (DSP) resources as well as on-chip data converters lend themselves perfectly to the needs of MKID readout hardware. The XCZU28DR \cite{rfsoc2}, for example, hosts 8 x 4.0 GSPS ADCs and 8 x 6.4 GSPS DACs, as well as 930,000 system logic cells and 4,272 DSP slices. During the comparison analysis of potential readout boards, Xilinx Inc. announced the release of their  ZCU111 board \cite{rfsoc1} which hosts the XCZU28DR chip. This announcement by Xilinx was followed by a number of related announcements from electronics companies stating that they would soon be releasing FPGA carrier boards housing these SoCs with high-speed interconnect to the chip among other features very useful for MKID readout requirements. As such, a number of these commercial systems were examined and compared against the more bespoke radio astronomy boards. Table \ref{possibleboards} shows a comparison of a representative sample of the systems which were examined. The third column, system cost, accounts for the additional components that would be required for each system. For some systems this includes DACs and ADCs, but the RF SoC systems do not require additional data converters. I/Q mixer boards are included for all systems, with an estimated cost of €5 K per board, and 1 board per 2,000 pixels. It should be noted that the stated costs are based on best estimates rounded to the nearest 1,000 Euro (as of December 2019). The fourth column in Table \ref{possibleboards}, quantifying the number of pixels each system is capable of reading out is not necessarily based on logic resources on the processing chip. For example, for the SKARAB board the 1,500 pixels is constrained by the bandwidth of the Peralex ADC which samples at a speed of 3 GSPS. Similarly, for the Xilinx VCU118 borad (which has huge logic capacity) we paired this board with a high-speed ADC with sampling speed of 4 GSPS. The cost of this ADC is included in the `System Cost' column.

\begin{table}[] 
\caption{FPGA Board Comparison. It should be noted that the costs for the systems are approximate, typically rounded to the nearest 1,000 Euro.} \label{possibleboards}
\begin{tabular}{p{3.2cm}p{1.4cm}p{1.8cm}p{1.2cm}p{2.2cm}p{3.4cm}}
\hline
 \textbf{Board} & \textbf{Unit Price(€)} & \textbf{System Cost (€)} & \textbf{Pixel Count} & \textbf{Cost/Pixel (€/pixel)} & \textbf{System-on-Chip (Y/N)} \\ \hline
 SKARAB &  11 k & 40 k & 1,500 & 26.67 & NO \\ \hline
 UniBoard II &  20 k & 40 k & 2,000 & 20.00 & NO \\ \hline
 AASL Board & 15 k & 35 k & 2,000 & 17.50 & NO \\ \hline
 Abaco VP880 & 15 k & 35 k & 2,000 & 17.50 & NO \\ \hline
 Xilinx VCU118 & 6 k & 25 k & 2,000 & 12.50 & NO \\ \hline
 \textbf{Xilinx ZCU111} & \textbf{9 k} & \textbf{19 k} & \textbf{4,000} & \textbf{4.75} & \textbf{YES} \\ \hline
 Abaco VP430 & 25 k & 67 k & 4,000 & 16.75 & YES \\ \hline
 VT AMC599 & 20 k & 28 k & 3,000 & 9.33 & NO \\ \hline
\end{tabular}
\end{table}

The Xilinx ZCU111 RF SoC clearly stands out from the rest of the systems in Table \ref{possibleboards}, at least based on cost per pixel. This cost of $\approx$€4.75 per pixel is assuming that 4,000 analogue probe tones can be generated, re-digitised, channelised and processed with the ZCU111. This assumption will be justified in Section \ref{sys_design}. In fact, there is enough bandwidth available from the 8 high-speed on-chip DACs and ADCs on the ZCU111 to readout over 8,000 MKID pixels. However, the logic resources on the chip/board would almost certainly be exhausted beyond 4,000 pixels. It was desirable to include an expected power consumption trade-off in Table \ref{possibleboards}, but obtaining or calculating the power demands for each system was not feasible during the current work. Instead, we simply stated whether each system employed an integrated system-on-chip, since this is expected to significantly affect power consumption.

As well as offering a very attractive cost per pixel, the ZCU111 also fulfills the other main aims of our next-generation readout system. 4,000 pixels per board is a factor x16 improvement on our current ROACH 1 system, and a x4 improvement on the state-of-the-art ROACH 2-based system. In fact, the eight high speed DACs could be used in pairs to synthesise 4 x 2,000 I/Q base-band tones with 2 MHz spacing over four parallel -2 to +2 GHz complex I/Q channels. Each of these parallel base-band waveforms could be up-mixed in frequency to the 4 - 8 GHz band with a 6 GHz LO, where each 4 - 8 GHz waveform would feed one of four separate feedlines. In this manner, up to 8,000 detectors could potentially be driven with just 4 feedlines (2,000 per feedline). In a similar manner, 4 pairs of the high speed ADCs could be used to re-digitise the 8,000 tones in I/Q. However, it is a near certainty that the logic resources available on the FPGA portion of the RFSoC will not be capable of processing this amount of data. For the near-term (until 2022), we will continue to work with programming the ZCU111 for just 4 DAC/ADC pairs, toward reading, channelising, and processing 4,000 resonators over two feedlines. But looking to the future with aims of utilising the full available bandwidth from this board, a brief discussion on combining the ZCU111 board with additional FPGAs is presented in section \ref{fmc_expand}.

As well as the reduced cost per pixel, and the increased number of pixels per board, the mass and volume are also significantly reduced compared to both ROACH 1 and ROACH 2 systems, simply due to the reduction in the number of boards needed for a given number of pixels and the integrated format of the data converters. Finally, the power consumption is also dramatically reduced. Estimates made by Xilinx claim that as much as half of the system power of a conventional FPGA/ADC/DAC board set is dissipated as heat in the Serialise/De-serialise (SERDES)) lines connecting the ADCs and DACs to the FPGA. This power dissipation is significantly reduced in the SoC format because chip to chip transmission of high-speed digital data is eliminated. Another major factor is the use of a more advanced CMOS node for the RFSoC vs. older FPGAs. Since at first sight the ZCU111 appears to meet all of our requirements, this board will form the basis of the remainder of this paper.

\section{System Design}\label{sys_design}
\subsection{Estimation of Usable Bandwidth on ZCU111}
Based on our requirements laid out earlier in this paper, this section outlines the broad structure of our DSP signal chain. Before any firmware or software was written for the new ZCU111-based system, an estimate was made for how much of the bandwidth (and thus how many pixels) could be generated, digitised, and processed. As a first order approximation, the logic resources of the ROACH 1 and ROACH 2 FPGAs were considered, and the number of MKIDs they could process was extrapolated based on the logic capacity of the RF SoC. The ROACH 1 based readout system used for the ARCONS array was capable of reading out 1024 pixels across 4 ROACH boards, achieving 256 pixels per board \cite{mcHugh1}. Moreover, the ROACH 2 based system used for the DARKNESS array could read out 10,240 pixels using 10 boards, achieving 1024 pixels per ROACH 2 board\cite{Meeker_2018}. The ROACH 1's Virtex 5 FPGA has 640 DSP slices, and 7,360 logic blocks \cite{virtex5_datasheet}, while the ROACH 2's Virtex 6 FPGA boasts resources of 2,016 DSP slices, and 476,160 logic blocks \cite{virtex6_datasheet}. In comparison, the Xilinx ZCU111 has 4,272 DSP slices and 930,000 logic cells \cite{rfsoc_dsp_slice}. Table \ref{fpga_resources} summarises this. 

\begin{table}[]
\caption{FPGA Resources Comparison} \label{fpga_resources}
\begin{tabular}{|l|l|l|ll}
\cline{1-3}
\textbf{Board} & \textbf{DSP Slices} & \textbf{Logic Blocks} &  &  \\ \cline{1-3}
ROACH 1        & 640                 & 7360                  &  &  \\ \cline{1-3}
ROACH 2        & 2016                & 476160                &  &  \\ \cline{1-3}
ZCU111         & 4272                & 930000                &  &  \\ \cline{1-3}
\end{tabular}
\end{table}

\begin{figure}[htbp]
\centerline{\includegraphics[width=90mm,height=60mm]{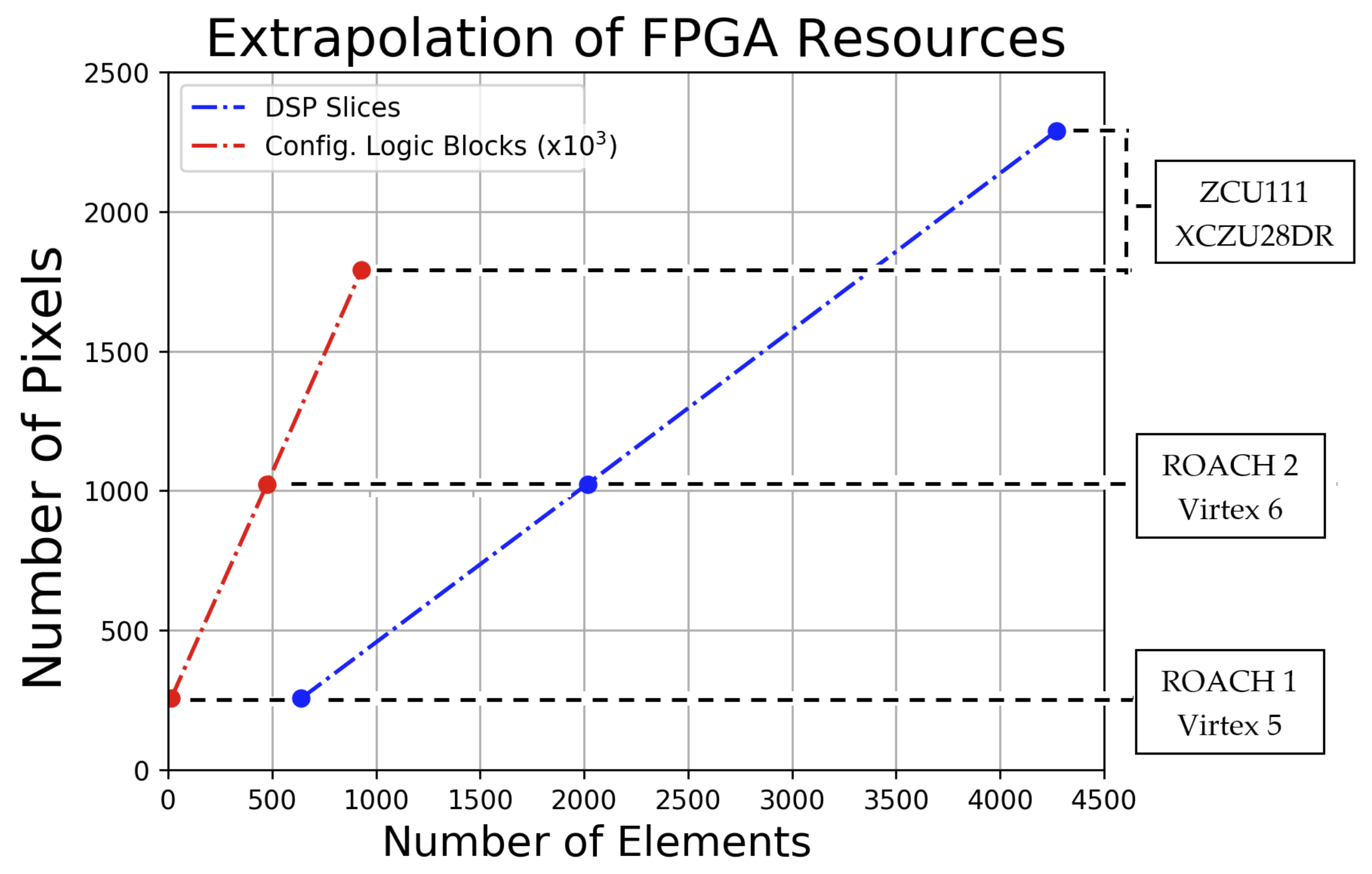}}
\caption{A very rough approximation of the number of MKID pixels that will be possible to read with the ZCU111 board.}
\label{fig:fpga_extrap}
\end{figure}

Fig.~\ref{fig:fpga_extrap} shows the extrapolation applied when both Configurable Logic Blocks (CLBs) and DSP Slices are considered. It should be noted that, to compare like with like, a time resolution of 1 $\mu$s was assumed, along with an assumed 2 MHz spacing between resonators in frequency space.

Now, Fig.~\ref{fig:fpga_extrap} seems to show that the ZCU111 will only be able to read~ 2,300 pixels if we base the extrapolation on DSP slices. In fact it only gets worse if we instead base the extrapolation on CLBs, resulting in only about 1,800 pixels; such a small multiplex (MUX) factor would make the development of the ZCU111 as an MKID readout hardly worthwhile. Fortunately, there are a number of factors that lead us to be confident that this number is an overly conservative estimate, and the actual MUX factor is at least double that shown in the extrapolations. These factors are as follows:
\begin{enumerate}
    \item{The number of MKIDs readout by the ROACH 2-based systems were limited by the sampling rates of the compatible data converters available at the time. In fact, less than half of the Virtex 6's logic resources were utilised in the DARKNESS ROACH 2 readout for channelisation \cite{strader1} and pulse detection.}
    \item{Our simple approach of extrapolating FPGA capabilities in the manner shown in Fig.~\ref{fig:fpga_extrap} is typically very conservative. DSP algorithms such as Polyphase Filter Banks (PFBs) are  not usually quite so linear, and if anything the resources scale in a more logarithmic manner. For example, the number of points in an FFT does increase linearly with bandwidth, but the number of taps in the PFB can in fact stay the same. Thus, there is no increase in the number of buffers and multiplies required, only the size of each buffer is extended \cite{PFB_Chang}. On-chip memory does tend to be linear though.}
    \item{We shall implement a more resource-efficient signal processing chain in firmware, significantly reducing the demands on FPGA logic resources. The firmware design is laid-out in Section \ref{course_chan}, and significantly reduces the processing required, compared to a twice-oversampled PFB}.
    \item{The DSP48E2 DSP slices which are specifically designed to carry the most intensive signal processing, have significantly improved since the Virtex 5 FPGA starndard. Most noticeable is the additional pre-adder in each DSP slice, which significantly improves performance in densely packed DSP designs and reduces the DSP slices required (for an FIR window filter for example, see below) by up to 50\% \cite{rfsoc_dsp_slice}. The configurable logic blocks (CLB's) also have extended capacity, and have increased routing and connectivity compared to Virtex CLBs. They also have additional control signals to enable superior register packing, resulting in overall higher device utilization. Table \ref{fpga_resource_comp} shows how the DSP slices and configurable logic blocks have improved over chip generations.}
    \item{Finally, the FPGA fabric clock on the ZCU111 will run at 500 MHz, twice the speed of the Virtex 6 fabric clock for the DARKNESS firmware design (250 MHz). This clock signal will be generated with the ZCU111's I2C programmable SI570 low-jitter differential oscillator (10 to 810 MHz).}
\end{enumerate}

\begin{table}[] 
\caption{DSP Slice and CLB Structure Comparison} \label{fpga_resource_comp}
\begin{tabular}{p{4.8cm}p{1.8cm}p{1.8cm}p{2.4cm}}
\hline
\textbf{Component / Device:} & \textbf{Virtex 5} & \textbf{Virtex 6} & \textbf{XCZU28DR} \\ \hline
\cellcolor[gray]{0.7}CLB \hspace{0.3cm}LUTs & \cellcolor[gray]{0.7}4 & \cellcolor[gray]{0.7}4 & \cellcolor[gray]{0.7}8 \\ \hline
\cellcolor[gray]{0.7}CLB \hspace{0.3cm}Flip-Flops & \cellcolor[gray]{0.7}4 & \cellcolor[gray]{0.7}8 & \cellcolor[gray]{0.7}16 \\ \hline
\cellcolor[gray]{0.9}DSP \hspace{0.3cm}Slice & \cellcolor[gray]{0.9}DSP48E & \cellcolor[gray]{0.9}DSP48E1 & \cellcolor[gray]{0.9}DSP48E2 \\ \hline
\cellcolor[gray]{0.9}\hspace{1.25cm}Multiplier &  \cellcolor[gray]{0.9}25 x 18 &  \cellcolor[gray]{0.9}25 x 18 & 27 × \cellcolor[gray]{0.9}18 \\ \hline
\cellcolor[gray]{0.9}\hspace{1.25cm}Accumulator & \cellcolor[gray]{0.9}48-bit & \cellcolor[gray]{0.9}48-bit & \cellcolor[gray]{0.9}48-bit \\ \hline
\end{tabular}
\end{table}
Memory should not be an issue as an additional memory can be added, such as a Hybrid Memory Cube (HMC) card, or High Bandwidth Memory (HBM) via the FPGA Mezzanine Card plus (FMC+) port on the ZCU111. 

\subsection{Coarse Channelisation}\label{course_chan}
\begin{figure}[htbp]
\centerline{\includegraphics[width=120mm,height=120mm]{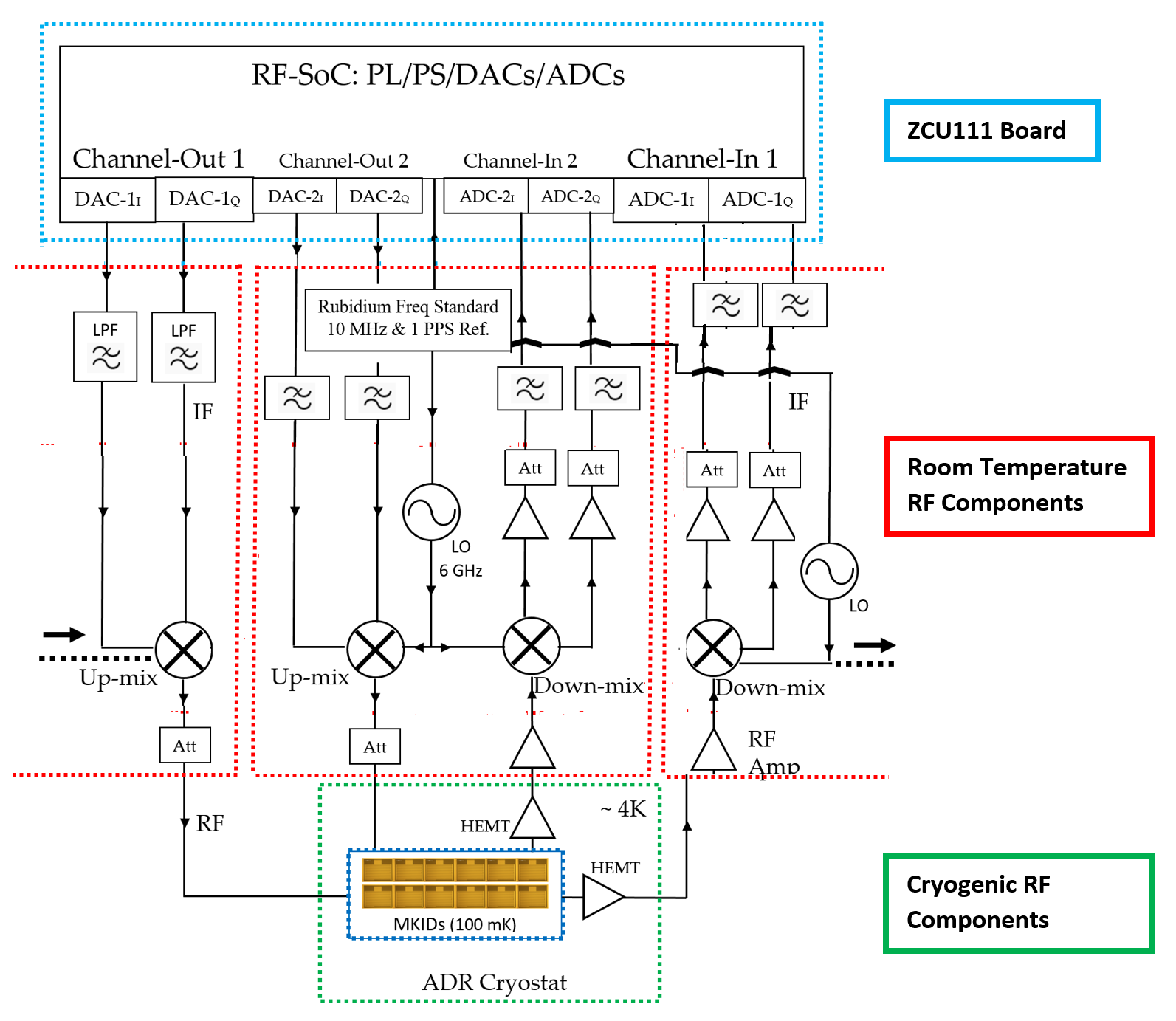}}
\caption{A block diagram showing the hardware components of our planned RF SoC-based readout system.The blue shell encompasses all electronic components housed on the ZCU111 board. The red shell is where the mixer components shall be housed: I and Q Inputs and Outputs, 6 GHz LO synthesiser, I/Q mixers, Ref Clock inputs, fixed-gain amplifiers and programmable attenuators. The green shell encloses the 3-stage cryogenic system, where the MKIDs will be  cooled to a base temperature of 100 mK.}
\label{fig:elec_schem1}
\end{figure}

Assuming that roughly 4,000 can be read and processed with the ZCU111 board, Fig.~\ref{fig:elec_schem1} shows how the design will be implemented in hardware, over two independent I/Q channels using 4 DACs and 4 ADCs. An I/Q up-mixer and down-mixer board with an LO frequency of 6 GHz will be required for each of the two channels, but a common Rubidium frequency standard will keep both channels and all components synchronised.

Considering the complex signal in just one of the two independent feedlines, a broadband complex `frequency-comb' (-2 to +2 GHz) will be generated by two DACs (I and Q). After these baseband signals are up-mixed to the 4 - 8 GHz and sent through the MKID array, they will subsequently be down-mixed back to -2 to +2 GHz band before being re-digitised by two of the ADCs at sample rates of 4 GSPS (again in I/Q format). We desire frequency bins with $\approx$1 MHz spacing, so we will apply a looped 4,096 point FFT to the time-stream data, resulting in FFT bins with just under 1 MHz spacing Fig.~\ref{fig:block_new1}. As previously mentioned in section \ref{SNR_req}, in order to ensure there are no scalloping losses in this coarse channelisation stage, while also trying to make our DSP as efficient as possible, we will implement a partially oversampled PFB, with oversampling ratio of 32/27. The result of this partially overlapping or staggering of data time-streams as they are fed into the buffers of the PFB causes a partial breakdown in the synchronisation of the sample timing in the sub-channels, which is discussed in the following section \ref{fine_chan}.

\subsection{Fine Channelisation and Phase-Rotation Correction}\label{fine_chan}
After performing the discrete FFT, the next step in signal processing is bin selection and refinement of those chosen FFT bins (often called fine channelization). The useful frequency bins (those sensitive to a probe tone) are selected, and a time-stream for each one is generated; we shall call these chosen bins `channels' now. Because of the partially oversampled approach used during the coarse channelisation stage, there will now be an observed oscillation in power in both the I and Q data time-streams for a tone that should be expected to be centred in a given channel. This is simply due to the slightly out-of-step sampling of each coarse FFT bin because of the fractional oversampling. However, since I/Q data will always be mutually 90$^o$ out of phase, by definition, the asynchronous effect will be observed in the complex signal as a rotation of phase in the I/Q plane. So, since it is the phase of each probe tone we will be monitoring in each channel, we must correct for this. A number of correction techniques are discussed in \cite{OverSampCorr_2017} for example.

\begin{figure}[htbp]
\centerline{\includegraphics[width=90mm,height=55mm]{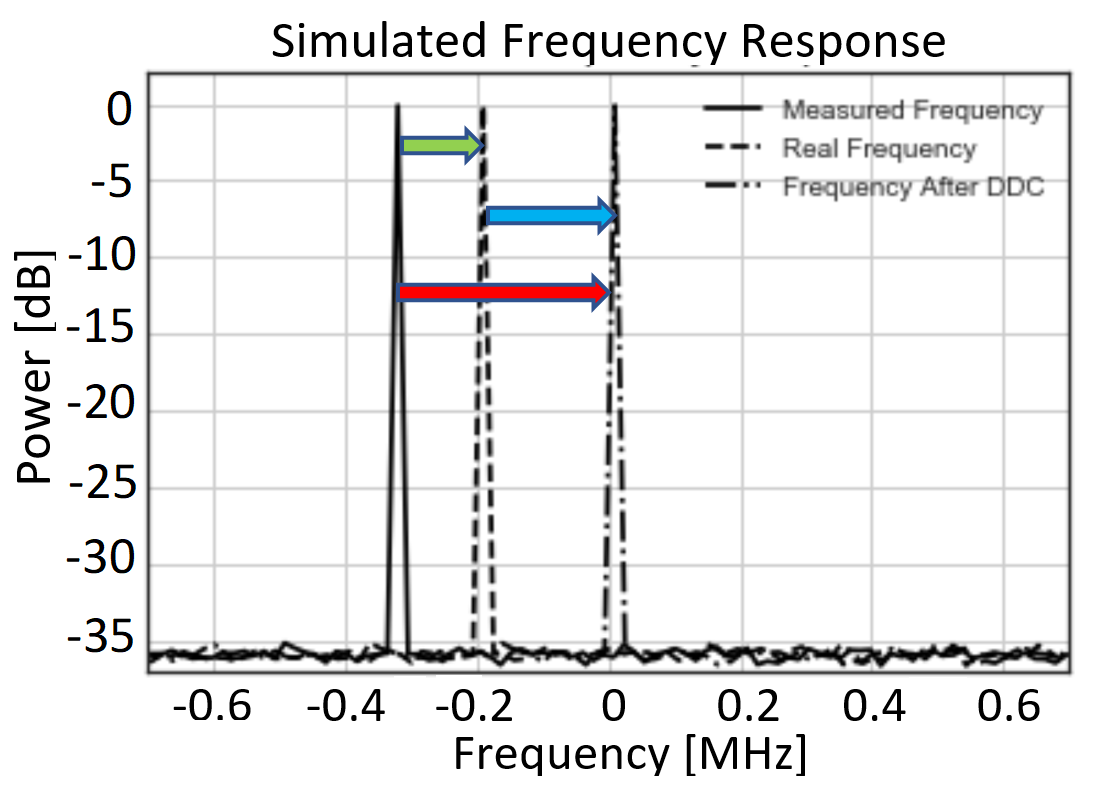}}
\caption{A simulated tone, if it were finely sampled within one of the 1 MHz frequency bins. The DDC must be given two frequencies to account for both the off-centre nature of the tone (200 kHz in this instance (blue arrow)), and the additional shift due to the non-maximally decimated sampling (133 kHz here (green arrow)).}
\label{fig:partial_OS_correction}
\end{figure}

Additionally, if the probe tone's frequency itself ($f_{tone}$) is not defined to be perfectly centered on it's corresponding channel, a similar oscillatory phase rotation effect will be observed in the I/Q plane. This oscillation will occur at a frequency equal to it's spectral distance from the channel centre. However, given that the tones are user-generated with precisely defined frequency values, these phase rotation frequencies are known apriori and can thus be corrected for. A Digital Down Conversion (DDC) step will be used for this correction, where each channel is digitally multiplied with a sinusoidal varying value. A benefit of using this DDC step is that we can define the frequencies of the DDC signals to also account for the phase rotation introduced by partial oversampling discussed above. The DDC signal for each frequency bin, sampled at $\approx$ 1 MHz is used to effectively shift each FFT bin to be centred on a particular tone, so that the measured power and phase remain constant over time. This shift is necessary, as we require each measured signal to be at 0 Hz when being monitored for evidence of a photon event. Once the additional phase rotation introduced by the 32/27 oversampling is calculated, it can simply be combined in the DDC calculation as shown in Eq.~\eqref{eq:res_shift_2} and illustrated in Fig.~\ref{fig:partial_OS_correction}.

\begin{figure}[htbp]
\centerline{\includegraphics[width=90mm,height=120mm]{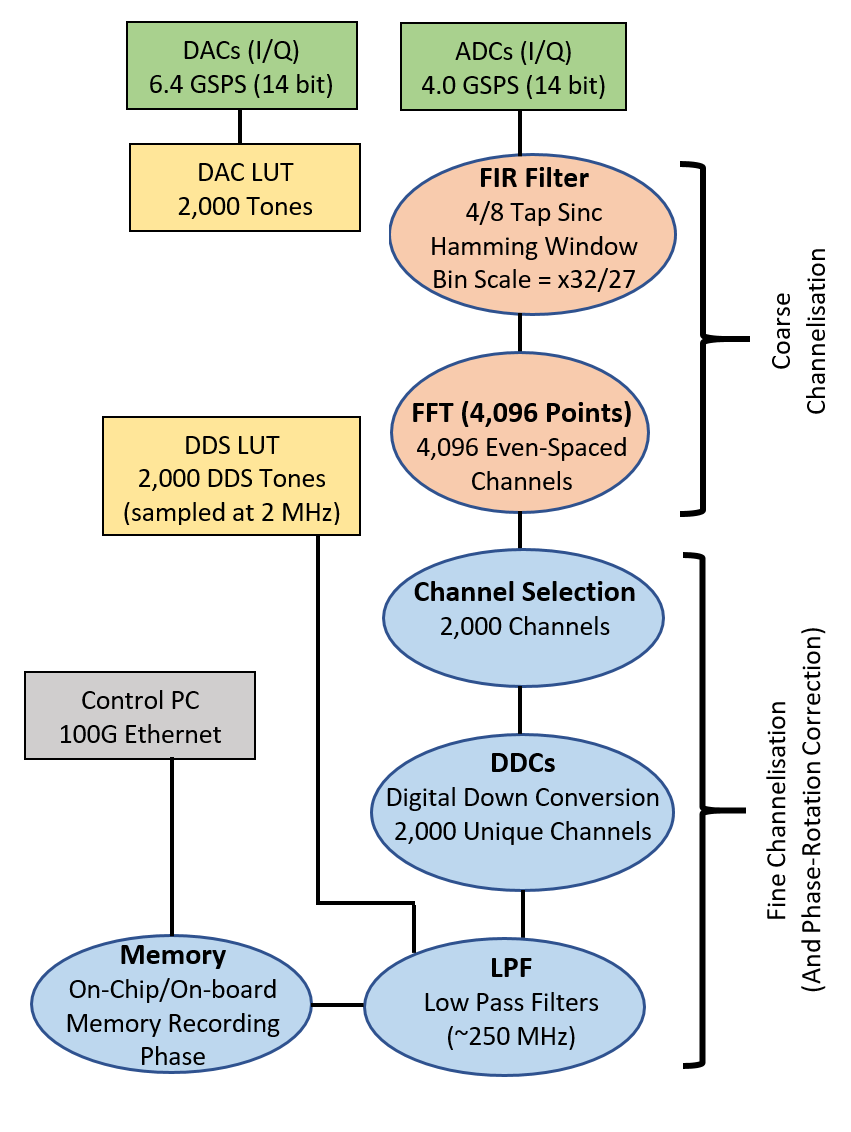}}
\caption{Block diagram showing a possible new DSP design, implementing a non-maximally decimated PFB for more efficient use of FPGA logic.}
\label{fig:block_new1}
\end{figure}

The DDC for each sub-channel will be achieved through a digital sine wave with frequency given by Eq.~\eqref{eq:res_shift_2}. Once the DDCs have been applied the time-stream from for each frequency bin with a probe tone will effectively be at DC. Now that we have corrected for the difference between the sampling of each probe tone and that tone's frequency, a low-pass filter (LPF) can be applied to each frerquency bin to also attenuate any other signals close to (or within) that bin. In a similar manner to that described by \cite{strader1}, a channel selection stage will then finally be applied, whereby any frequency bins with no tones within them will be discarded, and any frequency bins capturing more than one tone will be copied and treated in a similar manner to that just explained, with a DDC and LPF. In this manner we will go from having 4,096 evenly spaced, partially overlapped frequency bins, to roughly 2,000 uniquely spaced channels, where each unique channel will contain a single tone corresponding to a single MKID.  Fig.~\ref{fig:block_new1} shows a block diagram of the main steps in the DSP. Much of the design is based on scaling-up the firmware of other MKID instruments \cite{mcHugh1,strader1}, but uses the DDC in a novel way for correction of the unconventional sampling in the PFB.
\begin{equation}
\phi_f = f_{tone} - f_{0} - f_{shift}\label{eq:res_shift_2}
\end{equation}

\section{Alternative Implementations}
\subsection{Expanding to Secondary FPGA Carrier Boards} \label{fmc_expand}
Given the logic and memory resources appear to constrain the number pixels that are possible to read and process with the ZCU111 borad, a brief investigation was carried out into how the ZCU111's FMC+ port might be used to connect the RF SoC chip to external FPGAs. To minimise development costs and complexities, only COTS components were considered. The FMC+ port on the ZCU111 provides direct high-speed interconnect to the FPGA Programmable Logic (PL), and could serve as a link to external FPGAs for this additional processing capacity. The main disadvantage of this approach is the loss of one of the primary attractions of the RF SoC; namely the power-saving due to the removal of SERDES connections between FPGA and data converters. Although difficult to quantify the saving in power for the SoC design, a first order approximation shows a somewhat significant power saving simply based on the large heat load typically generated in the lines between FPGA and data converters in a standard non-SoC system..

The FMC+ port has 16 GTY transceivers (12 directly configured and 4 general I/O channels that can also be configured as GTY (Gigabit Transceiver - version 'Y'). Each of the ZCU111's GTY transceivers can, in principle, transfer up to 32.75 Gbps (gigabits per second) from chip-to-chip. So, in principle, an FMC+ port can transfer over 524 Gbps. Now, each of the ZCU111's 8 ADCs generates 4 GSPS, with each sample having 12 bits. This results in each ADC generating 48 gigabits of data every second.

Each GTY has two transceiver pins and can run at a maximum of 16.375 GHz PLL. The best approach is to run the internal GTY QPLL0 at 12 GHz (it can be clocked from 9.8 GHz to 16.375 GHz). Then, clocking the ADCs at 4 GHz, on each ADC duty cycle the 12 bit ADC sample will be split in 4, then serialised and sent across 2 GTY transceivers (4 Tx (transmit) pins in total, since there are 2 Tx pins per GTY).
So, we can send across 1 ADC sample (from 1 ADC), over 2 GTYs, per ADC duty cycle. Or in other words, on each GTY duty cycle we will transmit 1 bit on each Tx pin, and do this over 2 GTYs. Since the GTY duty cycle will be exactly 3 times faster than the ADC cycle, we can transmit 3 bits per GTY pin, per ADC clock cycle, for a total of 12 bits (or 1 sample) per ADC cycle.
Now, we can duplicate this 8 times, for the 8 ADCs, using up the full 16 available GTY channels (2 GTY for each ADC), and we don’t even have to clock the GTY at its maximum PLL speed. As such, all data from the 8 ADCs can be transferred continuously through this interface, and the bulk of the computationally intensive DSP can then be performed on the additional FPGA board.

\begin{figure}[htbp]
\centerline{\includegraphics[width=90mm,height=60mm]{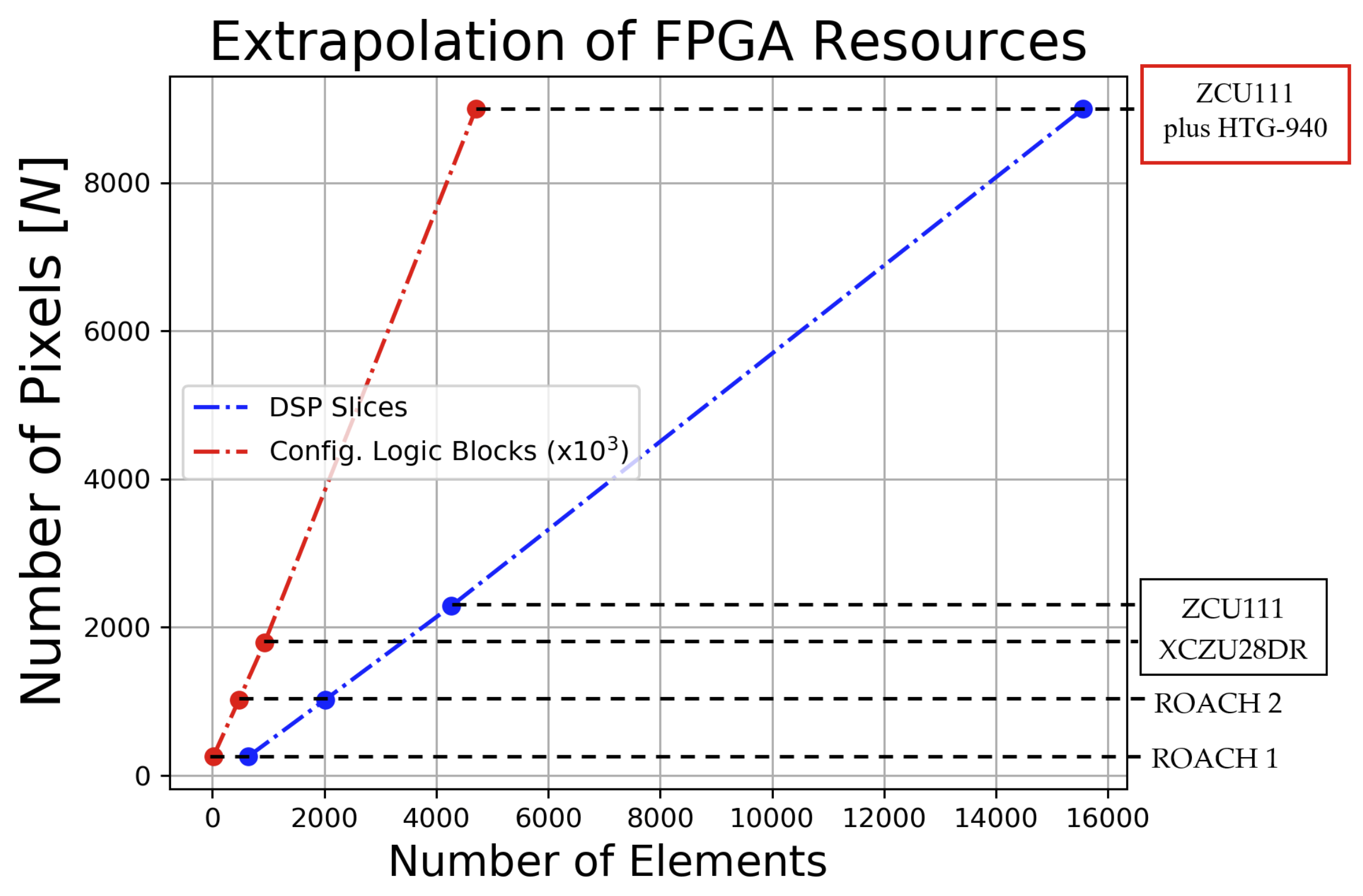}}
\caption{A very rough approximation of the number of MKID pixels that will be possible to readout if the ZCU111 board was to be combined with a secondary FPGA carrier board such as the HTG-940.}
\label{fig:fpga_extrap_FMC}
\end{figure}

The HTG-940 from Hitech Global, hosting the Virtex UltraScale+ and QUAD FMC+ ports would appear to be a sufficient solution for this format. Two of the four FMC+ ports may even allow two ZCU111 boards to be connected to it, since the UltraScale+ FPGA on the board has immense logic and DSP capacity (3.78 million system logic cells and 12,288 DSP Slices). The remaining two FMC+ ports could be used for HMC (hybrid memory cube) cards if needed. Fig.~\ref{fig:fpga_extrap_FMC} shows a similar plot to the extrapolation that was shown earlier, but now showing what would be possible if the ZCU111 was to be combined with the HTG-940 board, for example. The processing capacity of this system would approach 10,000 pixels, surpassing the 8,000 pixel bandwidth available. However, as shown in Table \ref{extra_boards}, the calculated cost per pixel is only marginally smaller with the addition of the HTG-940 board. Considering this, and the fact that exporting the ADC data off the ZCU111 chip to the external board will clearly consume more power due to dissipation, this option is not as attractive as first thought. In fact, it would seem to make more sense to simply use two ZCU111 boards to readout 8,000 resonators (4,000 per board), rather than a single ZCU111 connected an additional FPGA carrier board. The mass/volume footprint and cost per pixel are quite similar, the double ZCU111 approach would require less power per pixel to operate. Additionally, using two ZCU111 boards would allow the firmware and software to be cloned from one board to the next, avoiding the additional firmware and software development for the HTG-940 FPGA board. The HTG board hosts an FPGA from a different series, and the architecture is significantly different to that of the RF SoC chips.

\begin{table}[] 
\caption{Optical/Near-IR MKID Readout System Comparison. It should be noted that the costs for the systems are approximate, typically rounded to the nearest 1,000 Euro. A cost of $€$5 k has been included for each I/Q mixer board required by each system, assuming one mixer board per feedline running 2,000 pixels.} \label{extra_boards}
\begin{tabular}{p{3.2cm}p{1.4cm}p{1.8cm}p{1.2cm}p{2.2cm}p{3.4cm}}
\hline
 \textbf{Board} & \textbf{Unit Price(€)} & \textbf{System Cost (€)} & \textbf{Pixel Count} & \textbf{Cost/Pixel (€/pixel)} & \textbf{System-on-Chip (Y/N)} \\ \hline
 ROACH I & 5 k & 10 k & 250 & 40.00 & NO \\ \hline
 ROACH II & 5 k & 10 k & 1,000 & 10.00 & NO \\ \hline
 \textbf{Xilinx ZCU111} & \textbf{9 k} & \textbf{19 k} & \textbf{4,000} & \textbf{4.75} & \textbf{YES} \\ \hline
 \textbf{ZCU111 Plus HTG-940} & \textbf{17 k} & \textbf{37 k} & \textbf{8,000} & \textbf{4.63} & \textbf{NO} \\ \hline
\end{tabular}
\end{table}

\subsection{Expanding to Higher Frequency Bands}\label{generation_4}
Although the current RF-SoC-based readout system is only now being developed, it is never too early to start thinking to the future. In fact Xilinx Inc. have already released their so-called Gen 3 version of the RF SoC chips (our ZCU111 uses a Gen 2 version). An example of a carrier board that hosts one of these Gen 3 RF SoC chips is the Xilinx Zynq UltraScale+ RF SoC ZCU208 ES1 \cite{rfsoc_Gen3}. Similar in design to the board used for our current system, it again hosts 8 on-chip ADCs and 8 DACs, but the data converters are even faster. The ADCs clock at speeds up to 5 GSPS while the DACs run at speeds up to 10 GSPS. Both the ADCs and DACs are 14 bit, but most other specs on the board are essentially the same as the ZCU111. So, the main advantage this Gen 3 chip and board would offer the low temperature detector community is a higher frequency octave to work within, namely 5 - 10 GHz. The question then becomes, are there any sufficiently performing HEMTs on the market for this band (again pointing out that we are, possibly naively, demanding a $<$ 3 K noise temperature for our HEMTs). No doubt, there are exceptional HEMP amplifiers available for many frequency bands, from radio trough high-GHz to THz. However, we currently operate with the (possibly naive) assumption that our HEMT noise temperature will set a hard limit for our otipal/near-IR detectors. This assumed baseline may be naive because it is based on work carried out by McHugh et al. (2012) \cite{mcHugh1}. The HEMT was shown to be the highest contributor of noise for the first UVOIR MKID instrument, ARCONS \cite{arcons1}, where the HEMT noise was measured as roughly 40 dB above the ADC noise \cite{mcHugh1}. Whilst it was demonstrated that the available HEMTs at the time did appear to form a hard lower-limit of achievable noise, the quality of HEMTs has increased somewhat since 2012. Furthermore, the significantly broader bandwidths being demanded from our currently used ADCs may very well prove this assumption incorrect. If this proves to be true, then it would make sense to examine alternative HEMTs suitable for higher octaves. An analysis of measuring, and determining the highest contributing component of noise in our system will form the basis of a publication in the very-near future.

\begin{figure}[htbp]
\centerline{\includegraphics[width=90mm,height=60mm]{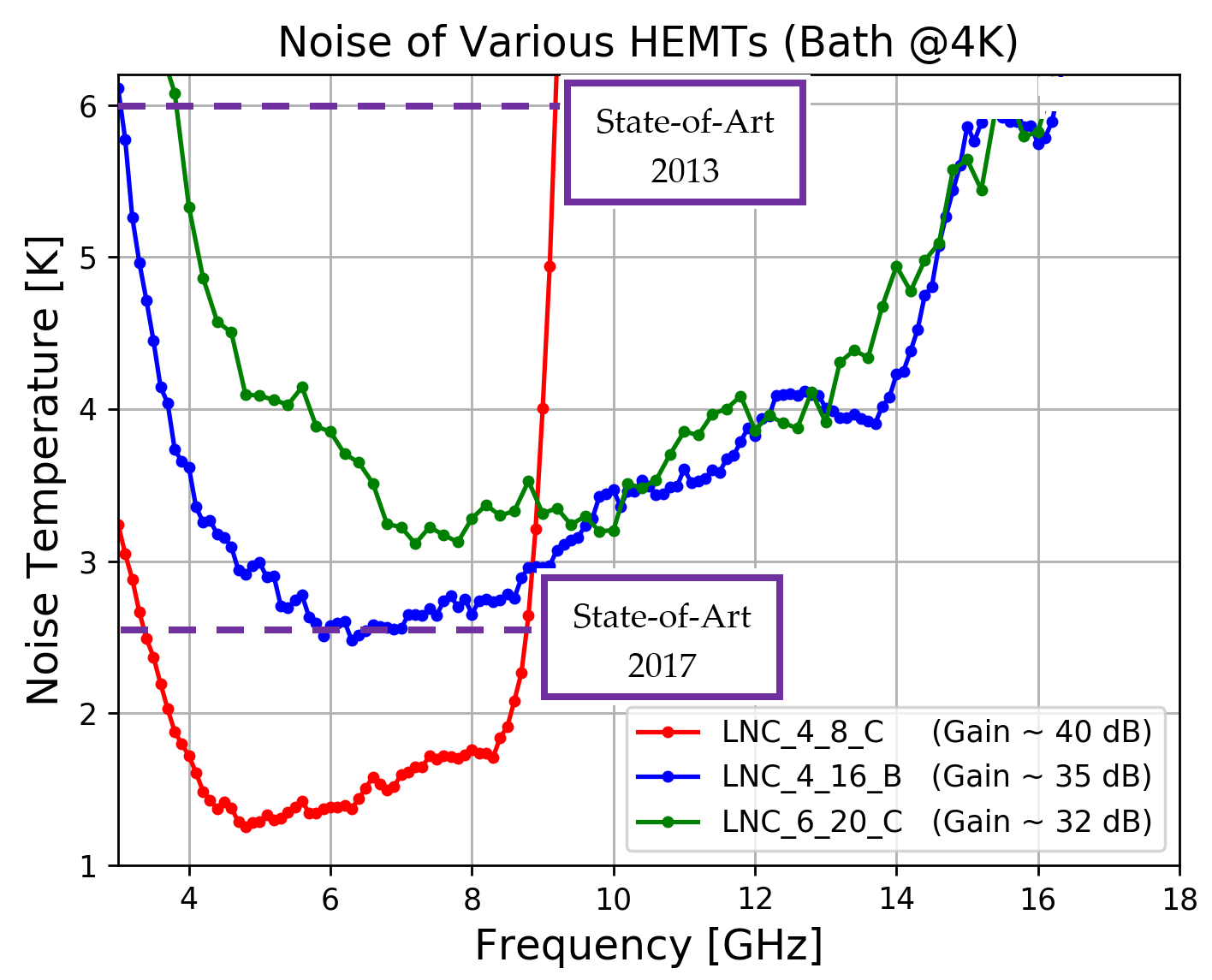}}
\caption{Noise temperature data for a number of LNF HEMT amplifiers. The gain levels of each of the three HEMTs shown are roughly 40 dB, 35 dB, and 32 dB across the frequency ranges shown. Data credit: Low Noise Factory, Sweden \cite{LNF_1,LNF_2,LNF_3}.}
\label{fig:hemt_1}
\end{figure}

For the moment, assuming that HEMT noise is our target baseline, it can be pointed out that significant progress has been made over the past 2 to 3 years in lowering the noise temperature of some cryogenic HEMTs that are operational over higher frequency ranges. Fig.~\ref{fig:hemt_1} shows the noise temperature data for three HEMTs that operate over three different frequency ranges. They are all manufactured by Low Noise Factory in Sweden, and the data is accurate as of July 20th, 2020.

With new boards such as the above-mentioned ZCU208 ES1 from Xilinx, the 5 - 10 GHz band should certainly be considered for future MKID instruments, especially seeing the excellent performance of the LNC4-16-B HEMT across this frequency band \cite{LNF_2}. Again though, the noise-floor of the on-chip components of these boards will need to characterised and compared to what is achievable with current HEMT technology, and acceptable noise levels. And assuming further increases in the frequency range of Xilinx RF SoC devices, the 6 - 12 GHz band should be kept in mind also. Assuming an equivalent spacing between resonators, these increased operational frequency ranges would correspond to a 25\% and a 50\% increase, respectively, in the number of pixels per feedline. As Fig.~\ref{fig:hemt_1} shows, even frequency ranges beyond 14 GHz \cite{LNF_3} are now capable of being kept well below the levels of HEMTs on previous instruments. 

The noise spectral density (NSD) contributed from the HEMT noise temperature can be calculated by Eq.~\eqref{eq:hemt} \cite{mcHugh1}.
\begin{equation}
NSD=\frac{k_{B}T_{n}}{P_{in}}\label{eq:hemt}
\end{equation}
The HEMT used in any cryogenic measurement work described here, has a noise temperature of roughly 2.3 - 2.7 K across the 4 - 8 GHz band. Assuming $T_n=2.7$ K, and aiming to drive our resonators at a power of around $P_{in} -75$ dBm, this results in a phase noise of -89.3 dBc/Hz when the signal is fully off-resonance (i.e., illuminated by a photon). However, since the resonator Q-factors can vary significantly, and this can effect the power with which they can be optimally driven, we should consider the case of $P_{in}\pm15$ dBm, namely -60 dBm and -90 dBm, which yields HEMT phase noise values of -104.3 dBc/Hz and 74.3 dBc/Hz, respectively. As such, we aim to keep all other sources of noise below the limit of -104.3 dBc/Hz. Detailed performance
of ZCU111, ZCU208, and HEMT amplifiers will continue as we implement systems.

\section{Summary of Conclusions}\label{conclusions}
With the aim of achieving a compact, low-power readout system for MKID arrays for optical/near-IR astronomy, this work describes a baseline design to be built around the Xilinx ZCU111 RF SoC system. Based on extrapolations shown and the application of more efficient channelisation methods, it is expected that this new system will be capable of reading out 4,000 resonators with 2 MHz spacing. Good agreement was shown between measurements and simulation for scalloping losses for the critically-sampled 4 tap PFB, with up to 6 dB of signal loss for tones at the edge of FFT frequency bins. A logic-efficient solution to prevent these losses was shown to be a non-maximally decimated PFB, whereby a flat response can be achieved across the full frequency bin without doubling-up on the PFB.
If more processing capacity is required, the ZCU111's on-board FMC+ connector was shown to provide enough high-speed interconnect to transfer the full data from all 8 on-chip ADCs of the ZCU111, to an external FPGA board via GTY transceivers. However, it was shown that, provided the ZCU111 itself can fully process 4,000 channels, then it is more sensible to simply use multiple ZCU111 boards to readout larger pixel numbers rather than exporting data to external FPGAs. It also was concluded that if TLS noise can not be kept below that of the HEMT, it will make most sense to extend our operational frequency range to a higher octave. Both the 5-10 and the 6-12 GHz octaves should be investigated further as a route to increasing the MUX factor.

\section{Acknowledgments}
This publication has emanated from research conducted with the financial support of Science Foundation Ireland under Grant number 15/IA/2880.

The authors would like to thank University of California Santa Barbara (UCSB), Mazin Labs, and the Collaboration for Astronomy Signal Processing and Electronics Research (CASPER) at University of California Berkley for their assistance. The MKID array in fig. 1 was fabricated by G.U. and is based on a design by Benjamin A. Mazin, UCSB. Thanks is also given to Xilinx and Cathal McCabe for donated FPGAs for the current ROACH 1 readout system and Vivado/ISE software under the Xilinx DUP programme. This work also benefited from remote discussions with an RF-SoC collaboration led by Gustavo Cancelo at Fermi National Accelerator Laboratory (Fermilab), Batavia, United States.

\textbf{Data Availability Statement:} The data that support the findings of this study are available from the corresponding author upon reasonable request.

\bibliography{Readout_RF_SoC_1}   
\bibliographystyle{spiejour}   

\listoffigures
\listoftables

\end{document}